\documentclass[iop,twocolappendix,numberedappendix,appendixfloats]{emulateapj}

\usepackage{natbib}
\usepackage[bookmarks,breaklinks]{hyperref}
\hypersetup{
	colorlinks,
	linktocpage,
	citecolor=blue,
	filecolor=magenta,
	linkcolor=blue,
	urlcolor=cyan
}
\usepackage{multirow}
\usepackage{longtable}
\usepackage{threeparttablex}
\usepackage{csquotes}
\usepackage{slantsc}
\usepackage{graphics}
\usepackage{adjustbox}

\newcommand{\OIII}{\mbox{[O\,\textsc{iii}]}}
\newcommand{\NII}{\mbox{[N\,\textsc{ii}]}}
\newcommand{\SII}{\mbox{[S\,\textsc{ii}]}}
\newcommand\tna{\,\tablenotemark{a}}
\newcommand\tnb{\,\tablenotemark{b}}
\newcounter{subfigure}

\shorttitle{SNIFS observations of 40 moderate-luminosity Type-2 AGNs}
\shortauthors{Rongxin Luo et al.}

\begin{document}
\title{Unraveling the complex structure of AGN-driven outflows:\\ V. Integral-field spectroscopy of 40 moderate-luminosity Type-2 AGNs}

\author{Rongxin Luo\altaffilmark{1,2}, Jong-Hak Woo\altaffilmark{1,3}, Marios Karouzos\altaffilmark{1}, Hyun-Jin Bae\altaffilmark{1}, 
Jaejin Shin\altaffilmark{1}, Nicholas McConnell\altaffilmark{4}, Hsin-Yi Shih\altaffilmark{5}, Yoo Jung Kim\altaffilmark{1}, Songyoun Park\altaffilmark{1}}
\altaffiltext{1}{Astronomy Program, Department of Physics and Astronomy, Seoul National University, Seoul 151-742, Republic of Korea}
\altaffiltext{2}{Department of Physics and Astronomy, University of Alabama in Huntsville, Huntsville, AL 35899, USA}
\altaffiltext{3}{Korea Astronomy and Space Science Institute, Daejeon 34055, Republic of Korea}
\altaffiltext{4}{Institute for Scientist and Engineer Educators, University of California, Santa Cruz, CA 95064, USA}
\altaffiltext{5}{University of Hawaii Maui College, 310 W Kaahumanu Ave, Kahului, HI 96732, USA}
\email{email: woo@astro.snu.ac.kr}

\begin{abstract}
There is an ongoing debate on whether feedback from active galactic nuclei (AGNs) can effectively regulate 
the star formation activities in their host galaxies. To investigate the feedback effect of AGN-driven outflows, 
we perform integral-field spectroscopic observations of 40 moderate-luminosity 
($10^{41.5} < L_{\mathrm{\OIII;cor}} < 10^{43.1}$ erg s$^{-1}$ ) Type-2 AGNs at z $<$ 0.1, 
which present strong outflow signatures in the integrated \OIII\ kinematics. Based on the radial profile of the normalized 
\OIII\ velocity dispersion by stellar velocity dispersion, we measure the kinematic outflow size and extend the kinematic 
outflow size-luminosity relation reported in \citet{Kang2018} into a wider luminosity range (over four orders of magnitude 
in \OIII\ luminosity). The shallow slope of the kinematic outflow size-luminosity relation indicates that while 
ionizing photons can reach out further, kinetic energy transfer is much less efficient due to various effects, demonstrating 
the importance of kinematical analysis in quantifying the outflow size and energetics. By comparing the outflow kinematics 
with the host galaxy properties, we find that AGNs with strong outflows have higher star formation rate and higher HI gas 
fraction than those AGNs with weak outflows. These results suggest that the current feedback from AGN-driven outflows 
do not instantaneously suppress or quench the star formation in the host galaxies while its effect is delayed.

\end{abstract}


\section{Introduction}
\label{sec:intro}
The energy output from active galactic nucleus (AGN) can be very high 
and if efficiently coupled with the ISM, it has the potential of significantly affecting 
the star formation in the host galaxies. AGN-driven outflows have been considered as an important channel to study 
the detailed feedback process (see \citealt{Elvis2000,Veilleux2005,Fabian2012} and \citealt{Heckman2014} for reviews). 
Many statistical works have observed the signatures of non-gravitational kinematics in the narrow line regions (NLRs) 
and indicate the prevalent existence of gaseous outflows among AGNs (e.g., \citealt{Nesvadba2008,Wang2011,Zhang2011,
Harrison2012,Mullaney2013,Bae2014,Genzel2014,Woo2016,Harrison2016,Wang2018,Rakshit2018,FoersterSchreiber2019,Leung2019}). 

Spatially resolved observations can be used to map the detailed kinematics of gaseous outflows and provide the measurements 
of outflow properties in different gas phases. Integral-field-spectroscopy (IFS) and radio interferometry observations have been 
used to traced the massive outflows of neutral and molecular gas as well as characterized their structures and kinematics in 
different objects and samples (e.g. \citealt{Cicone2012,Maiolino2012,Davies2014,Cazzoli2016,Rupke2017}). Outflows are 
found to be dominated by the molecular phase in term of mass \citep{Feruglio2010,Rupke2013,Cicone2014,GarciaBurillo2015,Fiore2017} 
and might be able to clean the gas content and quench star formation in the central region of galaxies \citep{Fluetsch2019}. 
Several studies have also shown enhanced star formation at the edge of and within AGN-driven outflows, 
suggesting a positive feedback effect \citep{Cresci2015a,Cresci2015,Maiolino2017,Carniani2016,Karouzos2016a,Shin2019}.

Adopting optical forbidden lines as a tracer of outflows, many spatially-resolved studies have begun to measure the 
detailed properties (e.g., geometry, kinematics, and energy) of ionized gas outflows in local AGNs and high-z QSOs 
(e.g.,\citealt{Sharp2010,StorchiBergmann2010,Liu2013,Liu2013a,Rupke2013,Harrison2014,Carniani2015,McElroy2015,
Husemann2016,Rupke2017,Revalski2018,Fischer2018,Freitas2018,Venturi2018,Circosta2018,Husemann2019,Davies2020,Scholtz2020}).
However, these studies mainly focused on the individual targets or small samples, there is a lack of systematic 
investigation on the ionized phase of AGN-driven outflows and their feedback effect based on the spatially 
resolved observations. 

To further characterize the properties of AGN-driven outflows and understand their feedback effect on the star formation 
processes in the host galaxies, \citet{Karouzos2016,Karouzos2016a}, \citet{Bae2017}, \citet{Kang2018} and \citet{Luo2019} 
have performed a series of spatially resolved studies of ionized gas outflows in a luminosity-limited sample of local Type-2 AGNs.
These targets are selected from a large sample of $\sim$ 39,000 Type-2 AGNs at z $<$ 0.3, which are used in our previous statistical 
study of outflow properties \citep{Woo2016}. This large sample focuses on the optically-selected AGNs, which are required 
to have high-quality Sloan Digital Sky Survey (SDSS) spectra and identified as AGN in the emission-line diagnostic diagrams.
Using the Gemini/GMOS-IFU data, \citet{Karouzos2016,Karouzos2016a} quantified the 
outflow properties and studied the excitation mechanism of ionized gas in six AGNs. They found the outflows are concentrated 
and detected circumnuclear star formation, suggesting the feedback effect of the outflow is limited and may not significantly 
impact the host galaxies. Based on the observations of Magellan/IMACS-IFU and VLT/VIMOS-IFU, \citet{Bae2017} extended 
the similar studies to a larger sample of 20 AGNs and found there is no clear evidence of instantaneous negative effect of 
the outflows. By combining the GMOS data in \citet{Harrison2014} and obtained from their own observations, \citet{Kang2018} 
performed a detailed measurement of the kinematic outflow size (see Section \ref{sec:size}) in 23 AGNs and found it is well 
correlated with the \OIII\ luminosity of AGNs. \citet{Luo2019} studied the kinematic properties of ionized gas in six AGNs with 
weak outflows and find that gas content in the host galaxies may play an important role to determine the occurrence of outflows.

In this paper, we present a spatially resolved study of 40 moderate-luminosity Type-2 AGNs at z $<$ 0.1, which have strong 
outflow signatures in the integrated \OIII\ kinematics. Using the data observed with the SuperNova Integral Field Spectrograph 
(SNIFS; \citet{Aldering2002,Lantz2004}), we measure the outflow size and kinematic properties and investigate their relations 
with the physical properties of AGN as well as the host galaxy, which can improve our understanding of the impact of AGN 
feedback. We describe the sample and observations in Section \ref{sec:sample}, and data reduction and analysis in 
Section \ref{sec:method}. We present the main results in Section \ref{sec:results}. Discussion and summary follow in 
Section \ref{sec:discussion} and \ref{sec:summary}.

\begin{deluxetable*}{c c c c c c c c}
	\tabletypesize{\footnotesize}
	\tablecolumns{8}
	\tablewidth{0pt}
	\tablecaption{Properties of the sample based on the SDSS spectra \label{tab:sample}}
	\tablehead{\colhead{ID}	& \colhead{V$_{\OIII}$}	& \colhead{$\sigma_{\OIII}$} & \colhead{$\log(\mbox{L}_{\OIII})$} &	
		\colhead{$\log(\mbox{L}_{\mbox{\OIII;cor}})$} & \colhead{m$_{r}$} & \colhead{b/a} & \colhead{Morphology Type} \\
		\colhead{ }	& \multicolumn{2}{c}{[km s$^{-1}$]}	& \multicolumn{2}{c}{[erg s$^{-1}$]} & \colhead{[AB]} & \colhead{ } & \colhead{ } \\
		\colhead{(1)} & \colhead{(2)} & \colhead{(3)} & \colhead{(4)} & \colhead{(5)} & \colhead{(6)}  & \colhead{(7)} & \colhead{(8)}}
	\startdata
	
	J1042+1808 &  86  & 456 & 39.8 & 41.8 & 15.25 & 0.84 &  tidal\tnb       \\                                                                  
	J1106+4530 &  81  & 429 & 40.0 & 42.0 & 14.76 & 0.90 &  spiral\tna      \\
	J0952+1937 & -204 & 477 & 40.1 & 41.6 & 14.27 & 0.84 &  spiral\tna      \\
	J1121+2825 & -283 & 292 & 40.1 & 42.3 & 17.09 & 0.82 &  elliptical\tnb  \\
	J2328+1446 & -216 & 422 & 40.2 & 41.9 & 16.91 & 0.93 &  elliptical\tnb  \\
	J1625+2228 & -83  & 417 & 40.2 & 42.5 & 15.34 & 0.35 &  spiral\tna      \\
	J0758+3747 &  70  & 380 & 40.2 & 41.7 & 12.94 & 0.60 &  elliptical\tna  \\
	J1236+1135 & -165 & 416 & 40.4 & 42.3 & 16.53 & 0.63 &  spiral\tna      \\
	J1040+3907 &  12  & 420 & 40.5 & 41.6 & 13.55 & 0.66 &  spiral\tna      \\
	J1208+5538 & -233 & 584 & 40.5 & 42.1 & 14.54 & 0.71 &  spiral\tna      \\
	J1019+5857 & -83  & 403 & 40.5 & 41.7 & 15.77 & 0.63 &  elliptical\tnb  \\
	J1339+0853 & -258 & 435 & 40.6 & 42.1 & 15.48 & 0.64 &  spiral\tna      \\
	J0144-0110 & -18  & 351 & 40.6 & 41.8 & 16.25 & 0.77 &  spiral\tnb      \\
	J1630+2434 & -164 & 442 & 40.6 & 42.1 & 15.32 & 0.42 &  spiral\tna      \\
	J1632+2622 & -73  & 379 & 40.6 & 42.0 & 14.67 & 0.52 &  spiral\tna      \\
	J1446+1122 & -46  & 371 & 40.6 & 42.0 & 14.88 & 0.99 &  spiral\tna      \\
	J1154+4555 & -66  & 404 & 40.6 & 41.8 & 15.34 & 0.66 &  spiral\tnb      \\
	J0130+1312 & -218 & 165 & 40.7 & 42.0 & 16.61 & 0.52 &  merger\tnb      \\
	J1301+2918 & -124 & 385 & 40.7 & 42.3 & 13.98 & 0.68 &  merger\tnb      \\
	J1200+0001 &  85  & 425 & 40.7 & 41.5 & 16.68 & 0.47 &  spiral\tnb      \\
	J1443+4759 & -215 & 266 & 40.9 & 42.4 & ...   & 0.75 &  elliptical\tnb  \\
	J1442+2201 & -249 & 438 & 40.9 & 42.0 & 16.24 & 0.96 &  spiral\tna      \\
	J1520+2846 & -37  & 438 & 40.9 & 42.2 & 16.59 & 0.70 &  elliptical\tnb  \\
	J1023+1251 & -59  & 369 & 41.0 & 42.1 & 15.63 & 0.59 &  spiral\tnb      \\
	J0836+4401 &  121 & 311 & 41.0 & 41.5 & 15.28 & 0.82 &  elliptical\tnb  \\
	J0753+1421 &  1   & 390 & 41.0 & 42.3 & 15.06 & 0.92 &  spiral\tna      \\
	J1344+5553 & -35  & 384 & 41.0 & 42.8 & 14.23 & 0.55 &  merger\tnb      \\
	J1550+2749 & -243 & 285 & 41.1 & 42.4 & 16.41 & 0.91 &  elliptical\tnb  \\
	J1327+1601 & -34  & 491 & 41.1 & 42.4 & 16.18 & 0.76 &  spiral\tna      \\
	J1037+5950 & -211 & 257 & 41.2 & 42.6 & 16.29 & 0.89 &  spiral\tna      \\
	J1654+1946 &  97  & 351 & 41.2 & 42.2 & 14.33 & 0.60 &  elliptical\tna  \\
	J1532+2333 & -65  & 359 & 41.2 & 42.3 & 16.29 & 0.59 &  elliptical\tnb  \\
	J1632+2349 &  29  & 386 & 41.3 & 43.1 & 15.35 & 0.95 &  elliptical\tnb  \\
	J1540+1049 & -78  & 443 & 41.3 & 42.6 & 17.01 & 0.73 &  uncertain\tnb   \\
	J1217+0346 & -92  & 398 & 41.3 & 42.3 & 15.78 & 0.93 &  spiral\tna      \\
	J1018+3613 &  155 & 484 & 41.4 & 42.8 & 14.67 & 0.90 &  tidal\tnb       \\
	J1213+5138 & -159 & 438 & 41.4 & 42.5 & ...   & 0.82 &  merger\tnb      \\
	J1434+0530 & -250 & 347 & 41.5 & 42.8 & 16.20 & 0.85 &  spiral\tnb      \\
	J1304+3615 & -222 & 354 & 41.5 & 42.8 & 15.43 & 0.77 &  elliptical\tna  \\
	J1147+5226 &  7   & 357 & 41.6 & 42.8 & 14.70 & 0.82 &  spiral\tnb
	
	\enddata
	\tablecomments{Col. 1: Target ID; Col. 2: \OIII\ velocity shift with respect to the systemic velocity; Col. 3: \OIII\ velocity dispersion; 
		Col. 4: Extinction-uncorrected \OIII\ luminosity; Col. 5: Extinction-corrected \OIII\ luminosity (see \citealt{Woo2016}); 
		Col. 6: \textit{r}-band magnitude from the SDSS photometry; Col. 7: Minor-to-major axis ratio; Col. 8: Morphology classification;}
	\tablenotetext{a}{Based on the classification from Galaxy Zoo project \citep{Lintott2011}.}
	\tablenotetext{b}{Based on our visual classification.}
\end{deluxetable*}

\section{Sample and observations}
\label{sec:sample}

\subsection{Sample selection}
\label{sec:selection}
\begin{figure}
	\centering
	\includegraphics[width=0.9\linewidth,angle=0]{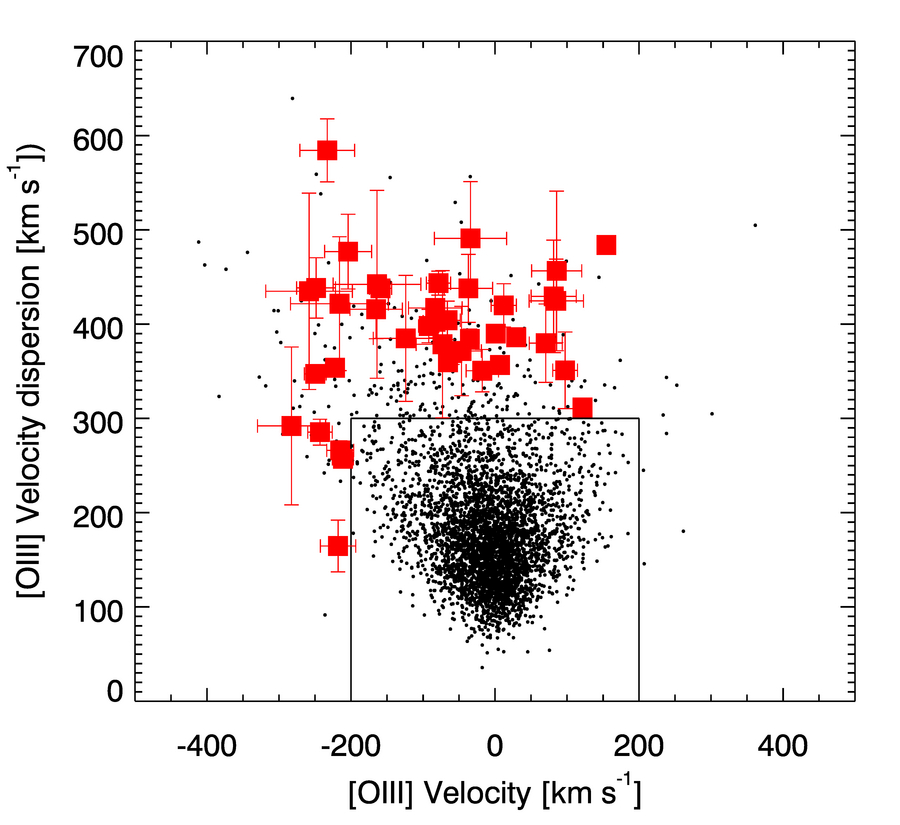}
	\caption{\OIII\ velocity versus velocity dispersion (VVD) diagram of the sample of 40 AGNs observed with SNIFS (red circles). 
		The \OIII\ velocity and velocity dispersion of these AGNs are measured from the SDSS spectra. The selection criteria of these 
		targets are indicated with the solid box. The luminosity-limited sample of local AGNs are denoted with black dots \citep{Woo2016}.} 
	\label{fig:sample}
\end{figure}

Based on the archival spectra of the SDSS Data Release 7, \citet{Woo2016} uniformly selected a large sample ($\sim$ 39,000) 
of Type-2 AGNs at z $<$ 0.3 and studied the properties and fraction of ionized gas outflows. 
Using a sample of 235,922 emission line galaxies from the MPA-JHU 
Catalog\footnote{\url{https://wwwmpa.mpa-garching.mpg.de/SDSS/}}, a parent sample of 106,971 Type-2 AGNs was
identified in the BPT diagram \citep{Baldwin1981,Veilleux1987} based on the classification scheme by \citet{Kauffmann2003}. 
Then, a sample of 38,948 Type-2 AGNs with well-defined emission line profiles was further selected based on two criteira, i.e., 
all targets were required to have a signal-to-noise ratio (S/N) $\ge$ 10 in the continuum, and with S/N $\ge$ 3 in the H$\beta$, 
\OIII$\lambda$5007, H$\alpha$, and \NII$\lambda$6584 emission lines, and the amplitude-to-noise (A/N) ratios were also required to be larger than 
5 for the \OIII$\lambda$5007 and H$\alpha$ emission lines. Note that AGNs with a very weak \OIII\ line are not included in this sample, suggesting
that the sample is biased against low-luminosity AGNs. 

We started our target selection from this sample for follow-up observations. Our selection criteria are listed as below. First, we limited the 
extinction-corrected \OIII\ luminosity as $L_{\mathrm{\OIII;cor}} > 10^{41.5}$ erg s$^{-1}$ and set a redshift cut of z $< 0.1$, which provide 
us a sample of luminous AGNs to perform spatially-resolved studies of the gas kinematic in the NLRs. \citet{Woo2016} adopted the Balmer 
decrement \citep[i.e., the H$\alpha$-to-H$\beta$ flux ratio of 2.86;][]{Netzer2009} to correct for the extinction of \OIII\ luminosity. In this case, 
the \OIII\ luminosity increases by an average factor of $\sim$ 7 and the $L_{\mathrm{\OIII;cor}} > 10^{41.5}$ erg s$^{-1}$ is corresponding 
to $L_{\mathrm{\OIII}} > 10^{40.7}$ erg s$^{-1}$. Following our previous studies \citep{Karouzos2016,Karouzos2016a,Bae2017,Kang2018}, 
we adopted the velocity shift and velocity dispersion measured from the \OIII\ emission line of SDSS spectra to trace the extreme kinematic 
signatures of ionized gas. We limited the \OIII\ velocity shift $|v_{\OIII}| > 200$ km s$^{-1}$ (with respect to the systemic velocity) and \OIII\ 
velocity dispersion as $\sigma_{\OIII} > 300$ km s$^{-1}$. The systemic velocity and stellar velocity dispersion ($\sigma_{\ast}$) were 
measured from stellar absorption lines in the SDSS spectrum \citep{Woo2016}. We finally selected 40 objects out of 902 Type-2 AGNs, 
which satisfy the selection criteria and the visibility of SNIFS. Note that our previous GMOS studies 
\citep{Karouzos2016,Karouzos2016a,Kang2018} focused on the AGNs with stronger outflow signatures 
(e.g., $\sigma_{\OIII} > 400$ km s$^{-1}$ and/or $L_{\mathrm{\OIII;cor}} > 10^{42}$ erg s$^{-1}$). 

As shown in Table \ref{tab:sample}, our SNIFS sample is dominated by spirals (21 objects) and ellipticals (12 objects) while 
the number of galaxies with merger or tidal features is five. Based on the comparison of larger samples of AGNs with and without outflow 
signatures, \citet{Luo2019} find that the presence of outflows may depend on the gas fraction in the host galaxies. Since the SNIFS
sample focuses on the AGNs with strong outflow signatures, the gas fraction of their host galaxies could be different from those of the
AGNs without outflow signatures. In addition, since the SNIFS sample is relatively small, it is difficult to directly compare with the un-biased
sample of the hard X-ray selected AGNs, which have a higher fraction of spiral/merger morphologies in their massive host galaxies \citep{Koss2010,Koss2011}.

The properties of the selected AGNs measured from the integrated SDSS spectra are presented in Table \ref{tab:sample}. 
As shown in Fig. \ref{fig:outflow_size1} and \ref{fig:outflow_size2}, the radial profiles of the ratio between \OIII\ velocity dispersion 
and stellar velocity dispersion become clearer as the extinction-uncorrected \OIII\ luminosity of these AGNs increase, thus we order 
them by this quantity in Table \ref{tab:sample} as well as in the following tables and figures. In Fig. \ref{fig:sample}, we present 
the \OIII\ velocity-velocity dispersion (VVD) diagram of the total luminosity-limited sample of the SDSS AGNs.

\subsection{Observations}
\label{sec:observations}
We used the SuperNova Integral Field Spectrograph (SNIFS) at the University of Hawaii's 2.2 m telescope (UH88) to observe the 
sample. SNIFS possesses a $6\farcs4\times6\farcs4$ field of view, which can cover 3-12 kpc scale at the redshifts of our targets. 
The field of view is subdivided into a grid of $15\times15$ spaxels, with a spaxel scale of $\sim 0\farcs43$ ($\sim$ 200-800 pc for 
our targets). The dual-channel spectrograph of SNIFS simultaneously covers the wavelength range of 3200-5200 \AA\
and 5100-10000 \AA, with the spectral resolution $R \sim 1000$ and 1300, respectively. For each target, the exposure 
time ranges from 1200 to 8400 seconds, which were determined based on the surface brightness of SDSS images. The 
observations were carried out under stable weather conditions (low wind, moderate humidity, and pressure values) in 
four observing runs in 2014 and 2015. The seeing values varied from $0\farcs9$ to $2\farcs1$ with a median of $1\farcs2$, 
corresponding to sub-kpc or kpc scale spatial resolution. The observation details are summarised in Table \ref{tab:observations}.

\begin{deluxetable*}{c c c c c c c}
\tabletypesize{\footnotesize}
\tablecolumns{7}
\tablewidth{0pt}
\tablecaption{Observational log of the tagets \label{tab:observations}}
\tablehead{\colhead{ID}	& \colhead{$\alpha_{2000}$} & \colhead{$\delta_{2000}$} & \colhead{z} & \colhead{Obs. Date} & 
\colhead{$t_{exp}$} & \colhead{Seeing} \\
\colhead{ }	& \colhead{[hh mm ss.ss]} & \colhead{[dd mm ss.ss]} & \colhead{ } &  \colhead{ }  &  \colhead{[sec]} &  \colhead{[\arcsec]} \\
\colhead{(1)} & \colhead{(2)} & \colhead{(3)} & \colhead{(4)} & \colhead{(5)} & \colhead{(6)}  & \colhead{(7)}}
\startdata
	
J1042+1808 & 10 42 22.36 & +18 08 06.72 & 0.0518 & 2015 Apr. 11        & 6000 & 1.1 \\
J1106+4530 & 11 06 23.94 & +45 30 39.96 & 0.0635 & 2015 Apr. 12        & 6000 & 1.2 \\
J0952+1937 & 09 52 59.03 & +19 37 55.41 & 0.0244 & 2015 Jan. 23        & 6000 & 1.5 \\
J1121+2825 & 11 21 12.99 & +28 25 10.92 & 0.0694 & 2015 Jan. 23        & 6000 & 1.8 \\
J2328+1446 & 23 28 06.06 & +14 46 24.96 & 0.0689 & 2014 Dec. 30        & 5400  & 1.0 \\
J1625+2228 & 16 25 07.18 & +22 28 53.76 & 0.0607 & 2015 Apr. 12        & 5400  & 1.0 \\
J0758+3747 & 07 58 28.11 & +37 47 11.84 & 0.0408 & 2014 Dec. 30        & 5400  & 1.1 \\
J1236+1135 & 12 36 34.51 & +11 35 34.08 & 0.0672 & 2015 Apr. 11,26     & 8400 & 1.1 \\
J1040+3907 & 10 40 00.57 & +39 07 19.99 & 0.0308 & 2015 Jan. 24,Apr. 25 & 7200 & 1.6 \\
J1208+5538 & 12 08 18.95 & +55 38 33.72 & 0.0513 & 2014 Dec. 30        & 5400  & 1.3 \\
J1019+5857 & 10 19 31.71 & +58 57 18.61 & 0.0420 & 2015 Jan. 25        & 6000 & 1.1 \\
J1339+0853 & 13 39 15.96 & +08 53 10.29 & 0.0662 & 2015 Apr. 12        & 6000 & 1.2 \\
J0144-0110 & 01 44 29.17 & -01 10 47.10 & 0.0604 & 2015 Jan. 24        & 6000 & 1.7 \\
J1630+2434 & 16 30 02.27 & +24 34 05.16 & 0.0619 & 2015 Apr. 12,13     & 6600 & 0.9 \\
J1632+2622 & 16 32 33.80 & +26 22 50.59 & 0.0587 & 2015 Apr. 25        & 5400  & 0.9 \\
J1446+1122 & 14 46 53.53 & +11 22 47.10 & 0.0519 & 2015 Apr. 26        & 3000  & 1.3 \\
J1154+4555 & 11 54 36.02 & +45 55 36.81 & 0.0431 & 2015 Jan. 24,Apr. 27 & 8400 & 1.3 \\
J0130+1312 & 01 30 37.76 & +13 12 51.99 & 0.0728 & 2015 Jan. 23,25     & 6000 & 1.2 \\
J1301+2918 & 13 01 25.27 & +29 18 49.43 & 0.0237 & 2015 Apr. 26        & 4200  & 1.2 \\
J1200+0001 & 12 00 37.69 & +00 01 26.04 & 0.0948 & 2015 Apr. 13        & 6000 & 0.9 \\
J1443+4759 & 14 43 54.85 & +47 59 24.36 & 0.0920 & 2015 Apr. 27        & 7200 & 1.4 \\
J1442+2201 & 14 42 32.21 & +22 01 16.46 & 0.0798 & 2015 Apr. 13        & 7200 & 0.9 \\
J1520+2846 & 15 20 14.44 & +28 46 11.28 & 0.0826 & 2015 Apr. 11        & 6000 & 1.5 \\
J1023+1251 & 10 23 06.98 & +12 51 00.36 & 0.0470 & 2015 Apr. 27        & 4800  & 1.0 \\
J0836+4401 & 08 36 37.84 & +44 01 09.58 & 0.0554 & 2015 Jan. 24,25     & 8400 & 1.3 \\
J0753+1421 & 07 53 28.28 & +14 21 40.97 & 0.0488 & 2015 Jan. 23        & 6000 & 1.5 \\
J1344+5553 & 13 44 42.15 & +55 53 13.81 & 0.0374 & 2015 Apr. 26        & 2400  & 1.3 \\
J1550+2749 & 15 50 01.60 & +27 49 00.48 & 0.0781 & 2015 Apr. 25        & 5400  & 0.9 \\
J1327+1601 & 13 27 34.65 & +16 01 11.28 & 0.0859 & 2015 Jan. 23        & 6000 & 2.1 \\
J1037+5950 & 10 37 41.50 & +59 50 50.02 & 0.0908 & 2015 Apr. 25        & 5400  & 1.5 \\
J1654+1946 & 16 54 30.73 & +19 46 15.53 & 0.0535 & 2015 Apr. 27        & 3600  & 1.4 \\
J1532+2333 & 15 32 22.32 & +23 33 24.98 & 0.0465 & 2015 Apr. 26        & 2400  & 1.3 \\
J1632+2349 & 16 32 13.44 & +23 49 09.73 & 0.0630 & 2015 Apr. 26        & 4800  & 1.3 \\
J1540+1049 & 15 40 55.43 & +10 49 09.48 & 0.0962 & 2015 Apr. 11,12,13  & 5400  & 1.2 \\
J1217+0346 & 12 17 41.99 & +03 46 31.01 & 0.0799 & 2015 Apr. 27        & 4800  & 1.0 \\
J1018+3613 & 10 18 33.63 & +36 13 26.40 & 0.0541 & 2014 Dec. 30        & 5400  & 1.1 \\
J1213+5138 & 12 13 03.35 & +51 38 54.97 & 0.0851 & 2015 Jan. 24        & 6000 & 1.7 \\
J1434+0530 & 14 34 37.87 & +05 30 16.16 & 0.0851 & 2015 Apr. 25        & 4800  & 0.9 \\
J1304+3615 & 13 04 22.17 & +36 15 43.12 & 0.0447 & 2015 Apr. 25        & 1200  & 1.5 \\
J1147+5226 & 11 47 21.60 & +52 26 58.78 & 0.0486 & 2015 Apr. 26        & 3600  & 1.2
	
\enddata
\tablecomments{Col. 1: Target ID; (2) Right ascension (J2000); (3) Declination (J2000); Col. 4: Redshift; 
Col. 5: Date of observation; Col. 6: Exposure time; Col. 7: Seeing size}
\end{deluxetable*}

\section{Data reduction and analysis}
\label{sec:method}

\subsection{Data reduction}
\label{sec:reduction}
All the data were pre-reduced at the observatory by using the standard pipeline for the Nearby Supernova Factory (SNfactory) project 
(see \citealt{Aldering2006} for details). Since our targets are AGNs, we adopted a special strategy for the data reduction and started 
from the intermediate data products of the SNfactory pipeline. These data products were overscan and bias subtracted, flat-field and cosmic 
ray corrected, wavelength calibrated, and telluric feature removed. First, we adopted the IDL wrapper SLA\_MOP\footnote{SLA\_MOP use the 
positional astronomy library SLALIB \citep{Wallace1994,Wallace2014} to calculate the atmospheric differential refraction as a function of 
wavelength, zenith angle, and weather conditions with a model atmosphere. See \url{http://homepage.physics.uiowa.edu/~haifu/idl.html} 
for details.} to calculate the atmospheric differential refraction offsets and used them to recenter the data cubes. Then we masked the 
AGN for each data cube and measured the corresponding sky background as the average flux in the rest region of the field of view.
The sky background was further subtracted from all the data cubes. During the processing of the SNfactory pipeline, all the data 
have been converted to units of erg s$^{-1}$ cm$^{-2}$ \AA$^{-1}$. We further performed the flux calibration by matching 
the flux in the SDSS spectra. The red and blue spectra were calibrated separately and then combined into a uniform wavelength range of 
3300-9700 \AA. In order to increase the S/N, we used a 2 spaxels $\times$ 2 spaxels moving box to smooth the data and obtained the final 
data cube with a format of $14\times14$ spaxels (corresponding to $6\farcs0\times6\farcs0$ field of view). The smoothing scheme did not 
change the initial spaxel scale of the data cube. With the spaxel scale of $\sim 0\farcs43$, we were able to Nyquist sample the seeing 
size (i.e. spatial resolution) and retain the spatial information of the targets.

\subsection{Data analysis}
\label{sec:analysis}
We followed the method as we previously adopted \citep{Karouzos2016,Bae2017,Kang2018,Luo2019} to perform the spectral 
analysis, which includes the subtraction of the stellar continuum and the extraction of the ionized gas kinematics. For each spaxel, 
we first fitted the continuum and measure the stellar velocity shift (with respect to the systemic velocity) and velocity dispersion 
by using the software pPXF \citep{Cappellari2004}. In this process, the continuum was modeled with 47 MILES simple-stellar 
population templates, which have solar metallicity and different ages ranging from 0.63 to 12.6 Gyr \citep{FalconBarroso2011}. 
In order to determine the systemic velocity of the host galaxy, we derived the spatially integrated spectra within the central 
3\arcsec\ spaxels and measured the velocity shift of the stellar absorption lines in them.

\begin{figure*}
	\centering
	\includegraphics[width=0.75\linewidth]{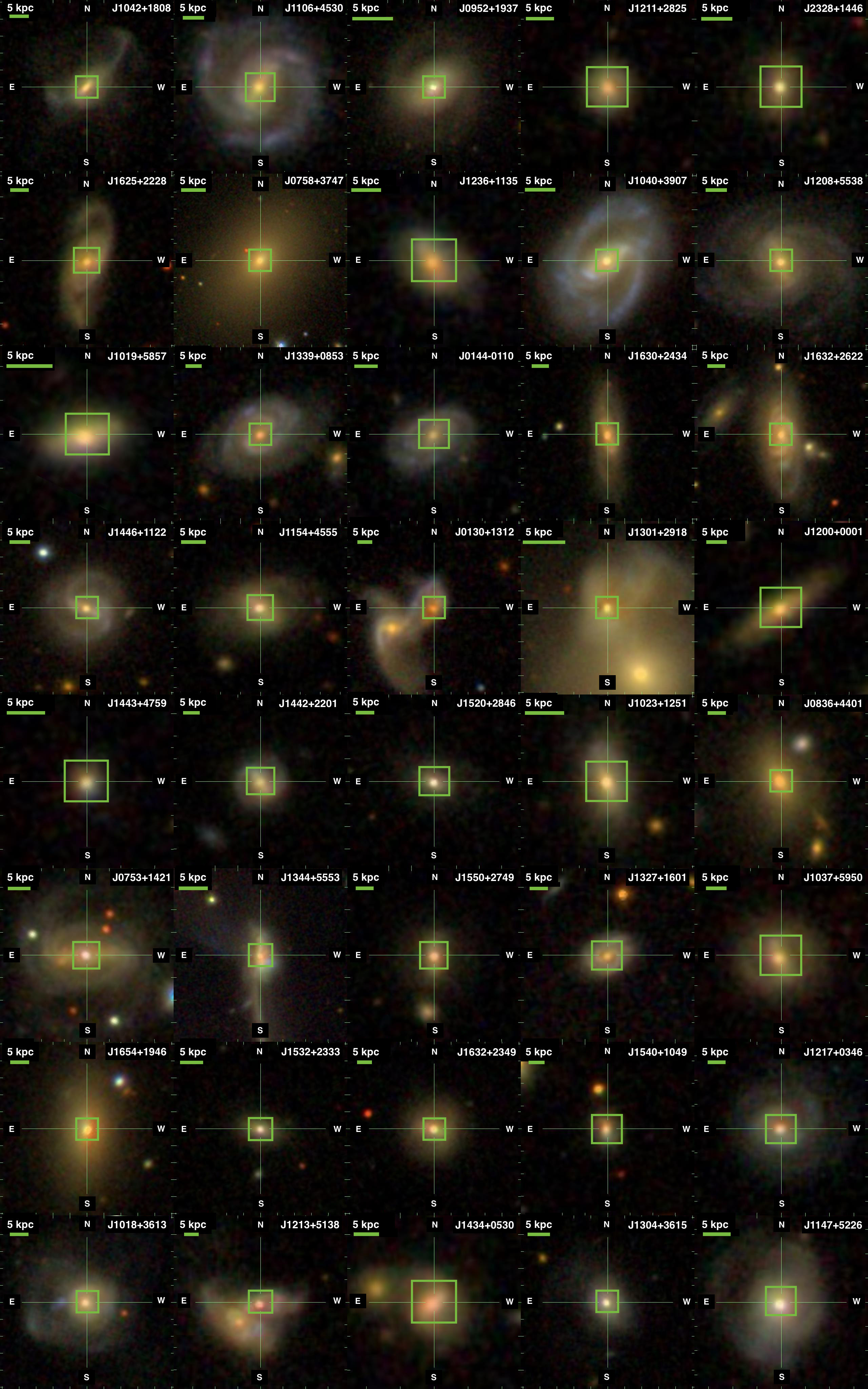}
	\caption{SDSS gri-composite images. The field of view of SNIFS IFU ($6\farcs4\times6\farcs4$) is shown with green boxes, 
		while the horizontal green bars show 5 kpc scale.}
	\label{fig:sdss}
\end{figure*}

\renewcommand{\thefigure}{\arabic{figure}\alph{subfigure}}
\setcounter{subfigure}{1}

\begin{figure*}
	\begin{adjustbox}{addcode={\begin{minipage}{\width}}{\caption{Examples of two dimensional maps ($6\farcs0\times6\farcs0$) 
						of eight targets, from left to right column: continuum (5030 -- 5170\AA) flux, \OIII\ flux, velocity and velocity dispersion, 
						\OIII/H$\beta$ flux ratio, BPT classification. The last column shows the spatially-resolved BPT diagrams 
						(\NII/H$\alpha$ vs \OIII/H$\beta$), with color-coded distance information. The horizontal white bars show 1 kpc scale. 
						Black contours denote continuum flux levels from 10\% to 90\% of the peak flux in steps of 10\% of the peak flux. The 
						dotted, dashed and solid lines show the demarcation lines defined by \citet{Kewley2001}, \citet{Kauffmann2003} and \citet{Cidfernandes2010}, respectively. \label{fig:onetarget1}
			}\end{minipage}},rotate=90,center}
		\includegraphics[width=1.25\linewidth]{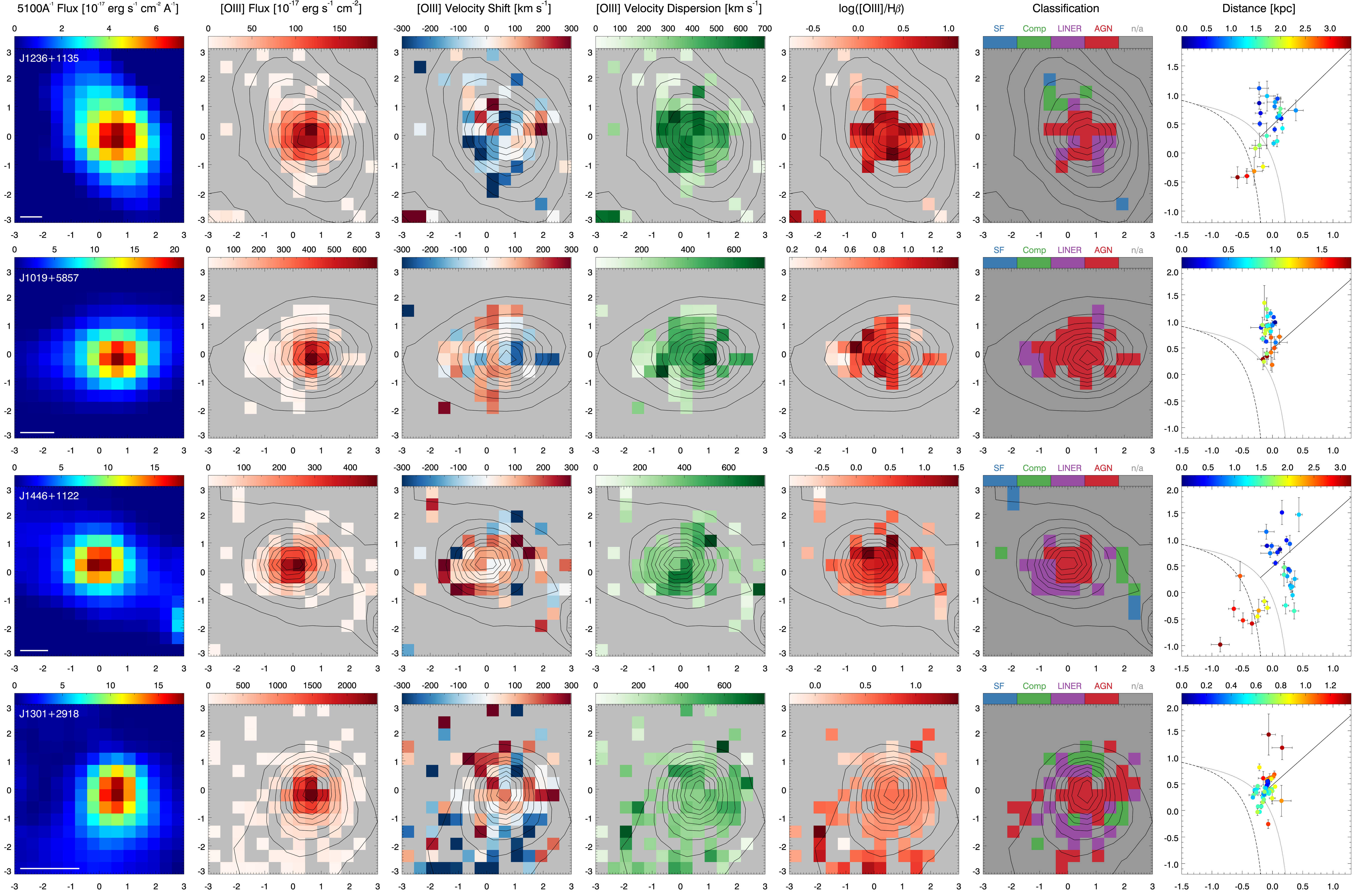}
	\end{adjustbox}
\end{figure*}

\addtocounter{figure}{-1}
\addtocounter{subfigure}{1}

\begin{figure*}
	\begin{adjustbox}{addcode={\begin{minipage}{\width}}{\caption{\label{fig:onetarget2}
			}\end{minipage}},rotate=90,center}
		\includegraphics[width=1.25\linewidth]{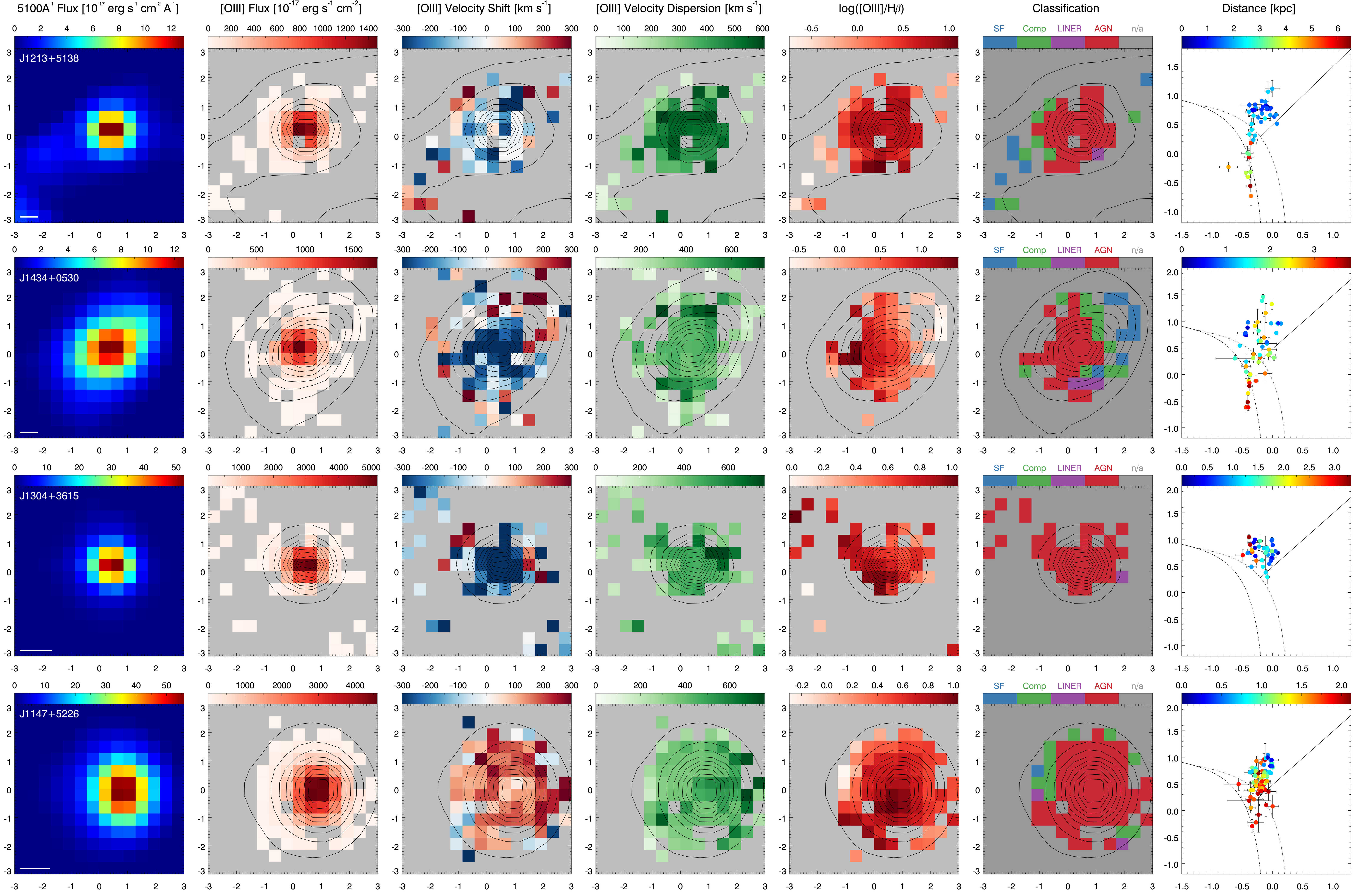}
	\end{adjustbox}
\end{figure*}

\renewcommand{\thefigure}{\arabic{figure}\alph{subfigure}}
\setcounter{subfigure}{1}

\begin{figure*}
	\begin{adjustbox}{addcode={\begin{minipage}{\width}}{\caption{Radial distributions of the ratio between \OIII\ velocity 
						dispersion and stellar velocity dispersion. The error-weighted mean (large blue circles) of $\sigma_{\OIII}/\sigma_{\ast}$ 
						in each distance bin and the measurements of each spaxel (dark gray points) are presented as a function of the radial 
						distance. The kinematic outflow size is defined when the mean of $\sigma_{\OIII}/\sigma_{\ast}$ becomes unity and 
						equals to 1 (red solid line). The range of the outflow boundary is represented by the shaded pink region. \label{fig:outflow_size1}
			}\end{minipage}},rotate=90,center}
		\includegraphics[width=1.25\linewidth]{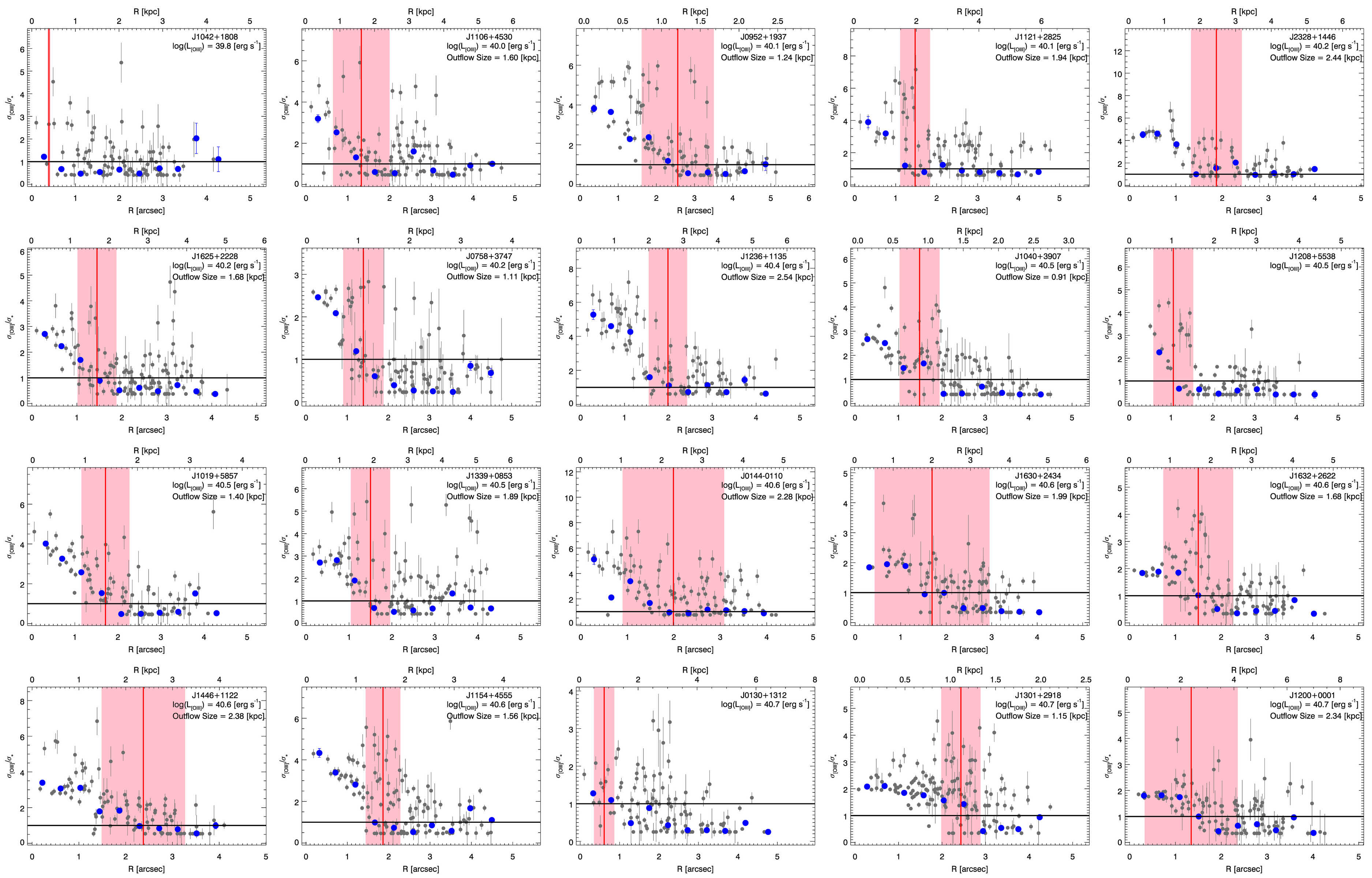}
	\end{adjustbox}
\end{figure*}

\addtocounter{figure}{-1}
\addtocounter{subfigure}{1}

\begin{figure*}
	\begin{adjustbox}{addcode={\begin{minipage}{\width}}{\caption{\label{fig:outflow_size2}
			}\end{minipage}},rotate=90,center}
		\includegraphics[width=1.25\linewidth]{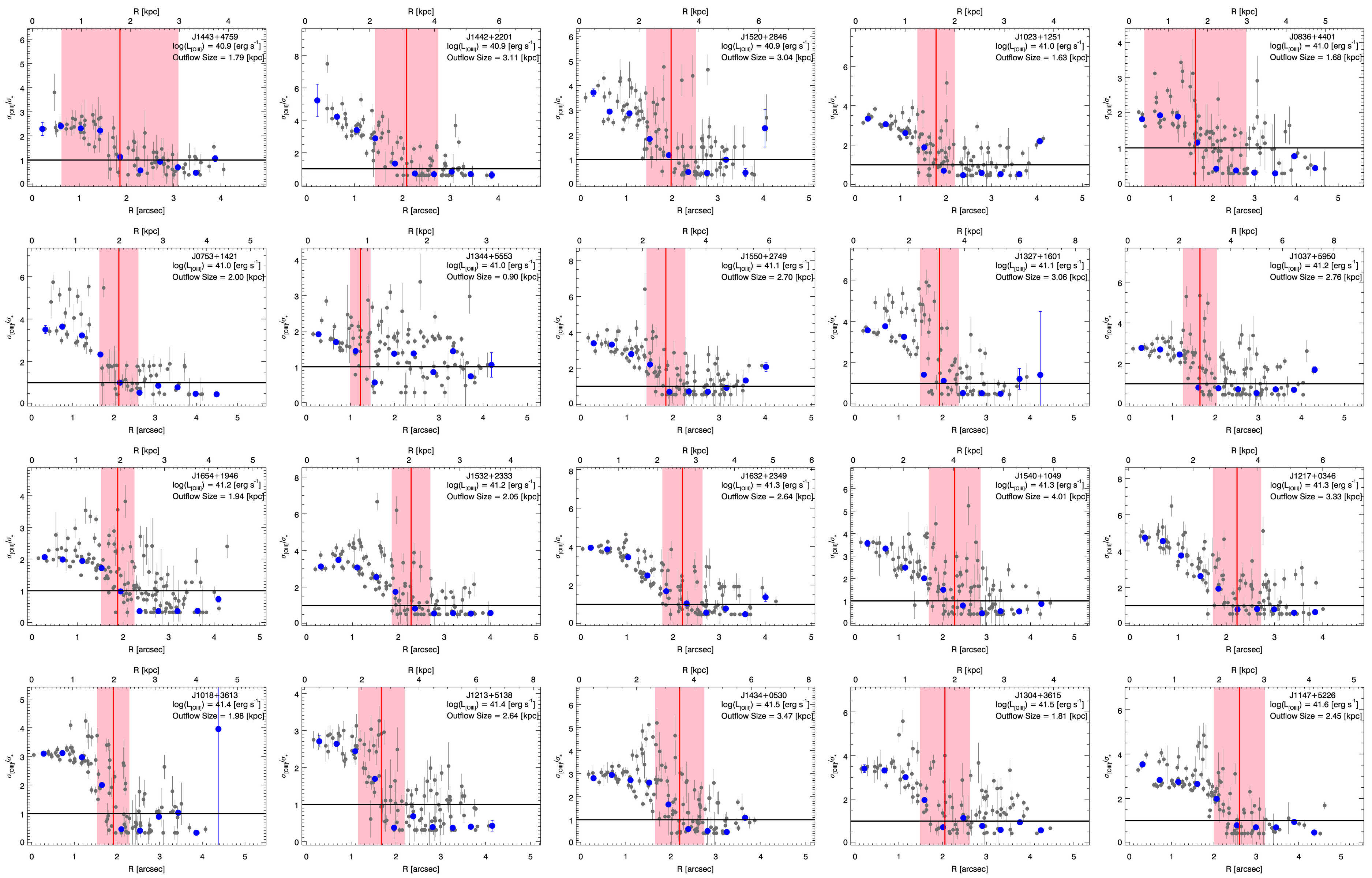}
	\end{adjustbox}
\end{figure*}

For the continuum-subtracted spectra, we used the IDL procedure \textit{MPFIT} \citep{Markwardt2009} to fit H$\beta$, \OIII, 
\NII, H$\alpha$, and \SII\ emission lines. Each line was fitted with up to two Gaussian components, including one narrow component 
and one broad wing component. We adopted two criteria to accept the fitting of the broad wing component: (1) Its peak amplitude 
is at least 2 times the continuum noise, which is determined as the standard deviation of the continuum-subtracted spectra at 
5100 -- 5200\AA. (2) The distance between the peaks of the two Gaussian components should be smaller than the sum of their 
widths ($\sigma$). For the H$\alpha$+\NII\ and \SII\ regions, to reduce the freedom degrees of the fitting, we assumed the same 
velocity shift and velocity dispersion for the doublets (\NII\ and \SII), while the velocity dispersion is in turn tied to the velocity 
dispersion of individual H$\alpha$ components. 

Based on the best-fitted line profiles, we measured the first moment $\lambda_{0}$ and second moment $\Delta\lambda^{2}$ for 
each emission line in each spaxel, which are defined as
\begin{equation}
\mathrm{\lambda_{0}}=\frac{\int\lambda f_{\lambda}d\lambda}{\int f_{\lambda}d\lambda},\;  \mathrm{\Delta\lambda^{2}}=\frac{\int\lambda^{2} f_{\lambda}d\lambda}{\int f_{\lambda}d\lambda}-\mathrm{\lambda_{0}}^{2}.
\label{eq:mom}
\end{equation}
Then we calculated the line flux, the velocity shift (with respect to the systemic velocity), and the velocity dispersion. 
We corrected the instrumental spectral resolution ($\sigma_{inst}\sim110$ km s$^{-1}$) by subtracting it in quadrature from 
the observed velocity dispersion. In addition, we performed Monte Carlo simulations to estimate the uncertainties of the 
measurements. By randomizing the flux using the flux error, we produced 100 mock spectra and fitted each of them. The 
standard deviation of the resulting distributions was adopted as the uncertainty. An iterative 4$\sigma$ clipping algorithm 
was used to exclude the bad fits in this process.

Based on the above spectral analysis, we obtained the two-dimensional maps of continuum flux and emission-line flux, 
ionized gas velocity and velocity dispersion of each target. The continuum flux is determined as the median value of the 
best-fitted continuum at 5030 -- 5170\AA. For the \OIII\ emission line, we employed an S/N limit of 3 (based on the peak S/N) 
to exclude spaxels with weak lines or bad measurements. Spaxels with lower S/N are masked as gray regions in the 
two-dimensional maps. 

\section{Results}
\label{sec:results} 
In Figure \ref{fig:sdss}, we present the SDSS gri-composite images of our sample and overlay the SNIFS field of view.
These targets have a variety of inclinations, with the minor-to-major axis ratio ranging from 0.35 to 0.99. We collect the 
information of morphology classification from the Galaxy Zoo project \citep{Lintott2011}. For the targets without classification 
in the Galaxy Zoo project, we perform visual morphological classification based on the SDSS gri-composite images. In summary, 
twenty-one galaxies are classified as spiral galaxies, which show spiral arms, bars, or ring structures. Six galaxies present 
clear tidal or merger signatures. There are also twelve elliptical galaxies and one galaxy (J1540+1049) without clear morphology 
classification. The detailed morphology classifications are presented in Table \ref{tab:sample}. 

\subsection{Emission Line Properties}
\label{sec:emission}
In Figure \ref{fig:onetarget1} and \ref{fig:onetarget2}, we present the two-dimensional maps of eight targets, which include the maps 
of continuum flux, \OIII\ flux, velocity and velocity dispersion, \OIII/H$\beta$ flux ratio, and BPT classification. The spatially resolved 
BPT diagrams are also shown. These targets are selected as examples of the low and high luminosity AGNs in our sample. 
Two-dimensional maps and spatially resolved BPT diagrams for other objects are shown in the appendix. In general, the spatial distribution 
of continuum flux shows a central-concentrated shape, indicating the bright nuclei of these AGNs. The \OIII\ flux is less extended than 
the continuum flux, while it presents an equally central-concentrated shape. Note that the less-extended distribution of \OIII\ 
flux is likely due to the observational constraints, since we require an S/N limit of 3 for the fitting of this emission line. In other words, 
the \OIII\ region could be more extended if we consider much weak \OIII\ fluxes.
In the central region, the [OIII] velocity appears blue- or redshifted and the [OIII] velocity dispersion increases, both signatures 
of AGN-driven outflows. We use the BPT diagram to investigate the ionization mechanisms. The criteria from \citet{Kewley2001} 
and \citet{Kauffmann2003} are adopted to classify the AGN, composite, and star-forming regions. For dividing Seyfert 
and LINER regions, we use the demarcation of \citet{Cidfernandes2010}. For all the targets, the central part is classified 
as Seyfert and/or LINER, indicating AGN-dominated photoionization. In fifteen targets, we also find signatures of both 
AGN and star-formation at the outer part of the emission regions, which is shown by the BPT classification of composite 
and star-forming regions. 
 
\subsection{Kinematic Outflow Size} 
\label{sec:size}
The information of spatially-resolved gas kinematics is very useful to quantify the outflow size. As shown in our previous 
studies \citep{Karouzos2016,Woo2016,Kang2017,Woo2017,Kang2018}, the \OIII\ velocity dispersion can trace the turbulent motion 
of ionized gas and present large values in the outflow regions. Following our previous strategy, we adopt the ratio between \OIII\ 
and stellar velocity dispersion ($\sigma_{\OIII}/\sigma_{\ast}$) to quantify the kinematic outflow size. Since $\sigma_{\ast}$ 
is considered as an indicator of the gravitational potential of the host galaxy, this ratio indicates the relative 
strength of the non-gravitational influence of AGN-driven outflows, and its radial change enables us to identify the edge of 
outflowing gas. The S/N of the stellar continuum of our SNIFS data is not high enough, thus we use $\sigma_{\ast}$ 
measured from the SDSS spectrum in the following analysis, which represents the global gravitational potential of the host galaxy.
Note that the stellar velocity dispersion measured from the SDSS spectra may be overestimated due to the rotational broadening. 
\citet{Kang2018} performed a consistency check to compare the stellar velocity dispersions measured from spatially-resolved GMOS 
spectra and those from the integrated spectra for 9 Type-2 AGNs. They found that the effect of rotational broadening is only a 
few percent.

\renewcommand{\thefigure}{\arabic{figure}}

\begin{figure}
	\centering
	\includegraphics[width=0.9\linewidth]{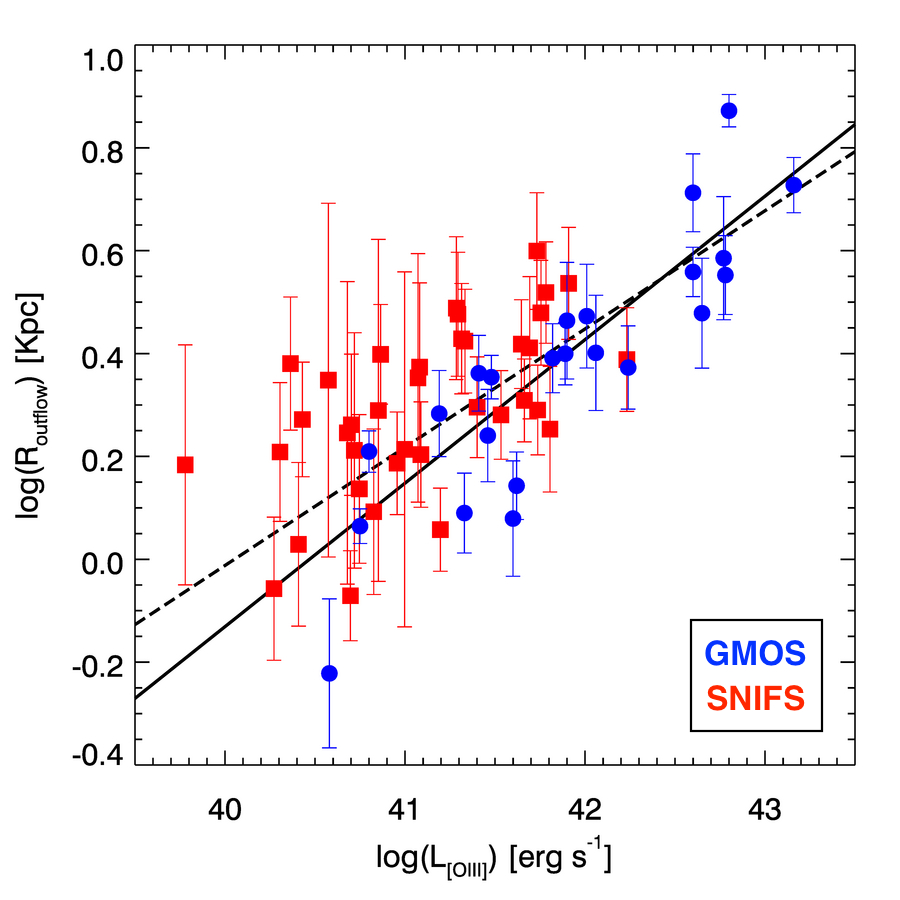}
	\caption{Correlation between the kinematic outflow size and the \OIII\ luminosity measured within the outflow size. 
		Red points show the measurements from our sample, while blue points present the results from \citet{Kang2018}. 
		The dashed line indicates the kinematic outflow size-luminosity relation (with the best-fitted slope of $0.230\pm0.024$) 
		obtained in this study, while the solid line presents the relation (with the best-fitted slope of $0.279\pm0.035$) 
		found in \citet{Kang2018} for comparison.}
	\label{fig:outflow1}
\end{figure} 

As shown in Figure \ref{fig:outflow_size1} and \ref{fig:outflow_size2}, for most targets in our sample, $\sigma_{\OIII}/\sigma_{\ast}$ 
from individual spaxels presents clearly a radial decrease and becomes unity at a large radial distance. This signature indicates that 
the ionized gas in the central region shows a strong non-gravitational component, while this effect becomes weak in the outer region 
of galaxies. To further quantify the radial behavior of the outflow kinematics, we define ten radial distance bins for each target and 
calculate the error-weighted mean of $\sigma_{\OIII}/\sigma_{\ast}$ in each bin. The kinematic outflow size is then determined when 
the mean of $\sigma_{\OIII}/\sigma_{\ast}$ becomes unity and equals to 1 (i.e. $\sigma_{\OIII}$ becomes comparable to $\sigma_{\ast}$). 
We also perform a seeing correction by subtracting the half-width-at-half-maximum of seeing from the kinematic outflow 
size in quadrature, which is 3\% on average. Considering the uncertainty of $\sigma_{\ast}$, we obtain the boundary of 
$\sigma_{\OIII}/\sigma_{\ast} = 1$ as well as the boundary of the kinematic outflow size. Note that the 2D velocity dispersion map 
typically shows asymmetric distribution, thus the average value in each radial bin has significant uncertainty. However, we tried to 
quantify a general trend as a function of radial distance. The range of this size boundary is adopted as 1$\sigma$ uncertainty 
of the kinematic outflow size. We exclude three targets (J0130+1312, J1042+1808, and J1208+5538) without a clear radial 
decrease of $\sigma_{\OIII}/\sigma_{\ast}$ and measure the kinematic outflow size for 37 AGNs, which ranges from 0.9 to 4 kpc 
with a median value of 2 kpc (see Table \ref{tab:properties} for details).

\begin{deluxetable*}{c c c c c c c c c c c}
\tabletypesize{\footnotesize}
\tablecolumns{11}
\tablewidth{0pt}
\tablecaption{Host galaxy and outflow properties \label{tab:properties}}
\tablehead{\colhead{ID}	& \colhead{sSFR} & \colhead{D$_{n}$(4000)} & \colhead{H$\delta_{A}$} & \colhead{M$_{\mbox{HI}}$/M$_{\ast}$} & 
\colhead{$\log(\mbox{M}_{\ast})$} & \colhead{$\sigma_{\ast}$} & \colhead{$\log(\mbox{L}_{\OIII})$} & 
\colhead{$\log(\mbox{V}_{bulk}/\mbox{M}_{\ast})$} & \colhead{$\log(\mbox{V}_{bulk}/\sigma_{\ast})$} & \colhead{Size$_{out}$} \\
\colhead{ }	& \colhead{[yr$^{-1}$]} & \colhead{ } & \colhead{ } & \colhead{ } & \colhead{[M$_{\odot}$]} & \colhead{[km s$^{-1}$]} & 
\colhead{[erg s$^{-1}$]} & \colhead{ } & \colhead{ } & \colhead{[kpc]} \\
\colhead{(1)} & \colhead{(2)} & \colhead{(3)} & \colhead{(4)} & \colhead{(5)} & \colhead{(6)}  & \colhead{(7)} & \colhead{(8)} & \colhead{(9)} &
\colhead{(10)} & \colhead{(11)}}
\startdata
	
J1042+1808 & -9.87  & 1.48 &  3.26 & ...   & 10.78 & 146.15 & ...   & ...  & ...  & ...  \\
J1106+4530 & -10.45 & 1.65 &  1.70 & 0.067 & 11.03 & 127.58 & 39.78 & 0.95 & 0.63 & 1.53$\pm$0.82 \\
J0952+1937 & ...    & 1.16 &  4.30 & 0.135 & 10.38 & 127.09 & 40.83 & 1.46 & 0.66 & 1.24$\pm$0.46 \\                                                          
J1121+2825 & -9.60  & 1.47 &  2.45 & 0.042 & 10.32 & 97.980 & 40.43 & 1.39 & 0.73 & 1.87$\pm$0.48 \\
J2328+1446 & -9.97  & 1.40 &  4.22 & 0.091 & 10.15 & 72.480 & 40.36 & 1.36 & 0.81 & 2.40$\pm$0.72 \\
J1625+2228 & -10.32 & 1.67 &  0.50 & 0.037 & 11.04 & 161.13 & 40.31 & 0.87 & 0.49 & 1.62$\pm$0.50 \\
J0758+3747 & ...    & 1.81 & -0.96 & 0.031 & 11.62 & 263.22 & 40.41 & 0.78 & 0.38 & 1.07$\pm$0.39 \\
J1236+1135 & -9.73  & 1.47 &  1.18 & 0.097 & 10.45 & 97.690 & 40.86 & 1.33 & 0.75 & 2.50$\pm$0.56 \\
J1040+3907 & ...    & 1.45 &  1.00 & 0.141 & 10.86 & 156.51 & 40.27 & 1.04 & 0.54 & 0.88$\pm$0.28 \\
J1208+5538 & -10.58 & 1.81 & -1.28 & 0.047 & 11.07 & 147.44 & ...   & ...  & ...  & ...  \\
J1019+5857 & -10.03 & 1.49 &  2.23 & 0.083 & 10.35 & 129.98 & 40.75 & 1.39 & 0.62 & 1.37$\pm$0.46 \\
J1339+0853 & -9.62  & 1.48 &  2.79 & 0.127 & 10.77 & 141.94 & 40.70 & 1.02 & 0.55 & 1.83$\pm$0.58 \\
J0144-0110 & ...    & 1.59 &  1.07 & 0.229 & 10.24 & 80.730 & 41.07 & 1.53 & 0.84 & 2.25$\pm$1.25 \\
J1630+2434 & -10.31 & 1.56 &  1.81 & 0.032 & 11.09 & 175.11 & 40.85 & 0.80 & 0.43 & 1.95$\pm$1.49 \\
J1632+2622 & -10.26 & 1.54 &  1.32 & 0.050 & 11.20 & 177.04 & 40.72 & 0.67 & 0.38 & 1.63$\pm$0.86 \\
J1446+1122 & -10.06 & 1.49 &  1.25 & 0.102 & 10.76 & 111.70 & 41.08 & 0.98 & 0.63 & 2.36$\pm$0.89 \\
J1154+4555 & -10.07 & 1.33 &  3.35 & 0.088 & 10.53 & 129.81 & 40.96 & 1.19 & 0.59 & 1.54$\pm$0.35 \\
J0130+1312 & -9.92  & 1.50 &  3.17 & 0.006 & 10.91 & 239.86 & ...   & ...  & ...  & ...  \\
J1301+2918 & ...    & 1.21 &  2.72 & 0.032 & 10.70 & 174.44 & 41.20 & 0.94 & 0.40 & 1.14$\pm$0.21 \\
J1200+0001 & -10.42 & 1.63 &  1.00 & 0.017 & 10.98 & 170.70 & 40.57 & 0.64 & 0.32 & 2.23$\pm$1.77 \\
J1443+4759 & -9.32  & 1.35 &  7.11 & ...   & 10.10 & 155.10 & 40.68 & 1.43 & 0.49 & 1.76$\pm$1.19 \\
J1442+2201 & -9.76  & 1.32 &  2.39 & 0.139 & 10.60 & 100.60 & 41.29 & 1.12 & 0.69 & 3.08$\pm$0.98 \\
J1520+2846 & -9.77  & 1.26 &  3.93 & 0.345 & 10.44 & 159.35 & 41.29 & 1.38 & 0.55 & 3.00$\pm$0.83 \\
J1023+1251 & -10.27 & 1.57 & -0.36 & 0.121 & 10.46 & 129.05 & 41.09 & 1.02 & 0.51 & 1.60$\pm$0.38 \\
J0836+4401 & ...    & 2.00 & -2.47 & 0.011 & 11.08 & 219.66 & 41.00 & 0.87 & 0.37 & 1.64$\pm$1.30 \\
J0753+1421 & -9.47  & 1.14 &  3.67 & 0.060 & 10.85 & 131.78 & 41.40 & 1.23 & 0.68 & 1.98$\pm$0.46 \\
J1344+5553 & ...    & 1.25 &  4.38 & 0.134 & 10.86 & 215.03 & 40.70 & 0.89 & 0.33 & 0.85$\pm$0.17 \\
J1550+2749 & -9.50  & 1.33 &  2.11 & 0.045 & 10.68 & 111.44 & 41.33 & 0.96 & 0.61 & 2.66$\pm$0.62 \\
J1327+1601 & -10.16 & 1.51 &  0.83 & ...   & 10.78 & 131.11 & 41.76 & 1.24 & 0.67 & 3.01$\pm$0.71 \\
J1037+5950 & -9.70  & 1.40 &  3.10 & 0.041 & 10.88 & 130.79 & 41.31 & 0.79 & 0.50 & 2.68$\pm$0.66 \\
J1654+1946 & -10.87 & 1.75 & -1.50 & 0.030 & 11.28 & 188.94 & 41.53 & 0.71 & 0.40 & 1.91$\pm$0.38 \\
J1532+2333 & -10.09 & 1.35 &  1.66 & 0.247 & 10.07 & 116.69 & 41.66 & 1.37 & 0.59 & 2.04$\pm$0.38 \\
J1632+2349 & -9.96  & 1.59 &  1.06 & 0.057 & 10.81 & 124.67 & 41.65 & 0.97 & 0.59 & 2.62$\pm$0.52 \\
J1540+1049 & -9.19  & 1.30 &  3.82 & ...   & 10.58 & 144.87 & 41.73 & 1.18 & 0.56 & 3.98$\pm$1.04 \\
J1217+0346 & -9.83  & 1.24 &  3.41 & 0.136 & 10.76 & 110.86 & 41.78 & 1.12 & 0.69 & 3.30$\pm$0.75 \\
J1018+3613 & ...    & 1.44 &  2.80 & 0.084 & 10.91 & 183.65 & 41.74 & 1.14 & 0.52 & 1.95$\pm$0.39 \\
J1213+5138 & -9.23  & 1.18 &  3.06 & ...   & 10.91 & 195.21 & 41.69 & 1.06 & 0.46 & 2.58$\pm$0.82 \\
J1434+0530 & -9.28  & 1.25 &  4.53 & 0.041 & 10.90 & 144.05 & 41.91 & 1.06 & 0.59 & 3.44$\pm$0.86 \\
J1304+3615 & -9.03  & 1.20 &  3.84 & 0.268 & 10.39 & 137.46 & 41.81 & 1.50 & 0.63 & 1.79$\pm$0.50 \\
J1147+5226 & -9.61  & 1.24 &  4.75 & 0.310 & 10.69 & 142.94 & 42.23 & 1.03 & 0.53 & 2.45$\pm$0.57

\enddata
\tablecomments{Col. 1: Target ID; Col. 2: Specific star formation rate; 
Col. 3: D$_{n}$(4000); Col. 4: H$\delta_{A}$; Col. 5: HI gas fraction; Col. 6: Stellar mass; Col. 7: Stellar velocity dispersion; 
Col. 8 Extinction-uncorrected \OIII\ luminosity measured within the outflow size: Col. 9: Bulk velocity normalized by stellar mass; 
Col. 10: Bulk velocity normalized by stellar velocity dispersion; Col. 11: Seeing-corrected outflow size}
\end{deluxetable*}

\section{Discussion}
\label{sec:discussion}

\subsection{Relation between Outflow Size and AGN Parameters}

In order to understand the driving mechanism of the outflows, we examine the relation between outflow properties and AGN parameters. 
We first compare the kinematic outflow size with the \OIII\ luminosity measured within the outflow boundary to be consistent with our previous studies 
performed \citep{Karouzos2016,Bae2017, Kang2018}. As shown in Figure \ref{fig:outflow1}, 
the kinematic outflow size is well correlated with the \OIII\ luminosity, as reported by \citet{Kang2018}. Based on the GMOS 
observation of 23 Type-2 AGNs over 3 orders of magnitude in \OIII\ luminosity, \citet{Kang2018} has presented the kinematic outflow 
size-luminosity relation with the best-fit slope of $0.279\pm0.035$. By combining their sample with our size measurements, 
we can now extend the correlation with a large sample of 60 Type-2 AGNs over a broad \OIII\ luminosity range. 
We perform a forward regression using the MPFITEXY routine \citep{Williams2010} and obtain the best-fit relation:
\begin{equation}
\rm log (\frac{R_{\mathrm{out}}}{\mathrm{Kpc}})=(0.230\pm0.024) \times \rm log (\frac{L_{\mathrm{\OIII}}}{10^{42} \mathrm{erg s}^{-1}}) + (0.448\pm0.020).
\label{eq:size_lum}
\end{equation}
This relation is generally consistent with that obtained by \citet{Kang2018}, which has a slightly higher slope with a smaller sample. 
We also compared the kinetic outflow size with black hole mass and Eddington ratio. However, we find no significant correlation.
The correlation between the kinematic outflow size and \OIII\ luminosity indicates that the AGNs with higher luminosity produce 
outflows in a larger scale, leading to more significant influence on the global kinematics of the host galaxy.

Based on the flux distribution of the \OIII\ emission line, the NLR size of AGNs has been quantified using different methods and 
various slopes of the NLR size-luminosity relation have been reported in the literature. \citep{Bennert2002}, \citet{Schmitt2003} and 
\citet{StorchiBergmann2018} used the HST narrow-band images to measure the maximum detectable size of emitting regions in several 
AGN samples and obtained the best-fit slopes of $0.52\pm0.06$, $0.33\pm0.04$, and $0.51\pm0.03$, respectively. \citet{Greene2011} 
performed a similar analysis on the long-slit data of 15 radio-quiet obscured quasars, reporting a best-fit slope of $0.22\pm0.04$, while 
\citet{Fischer2018} presented a best-fit slope of $0.42\pm0.03$ by combining the HST STIS long-slit data of 12 Type-2 quasars and the 
NLR size measurements from \citet{Schmitt2003}. \citet{Husemann2014} and \citet{Bae2017} used an alternative definition of the NLR 
size and measured the flux-weighted effective radius from the IFU data of local AGN samples, reporting best-fit slopes of $0.44\pm0.06$ 
and$0.41\pm0.02$, respectively. Other studies adopted a fixed limit of the \OIII\ intrinsic surface brightness to quantify the NLR size of 
AGNs. From the IFU and long-slit observations of radio-quiet obscured quasars, \citet{Liu2013} obtained a best-fit slope of the NLR
size-luminosity relation as $0.25\pm0.02$, and \citet{Hainline2013} also reported a consistent result. The slope of kinematic outflow
size-luminosity relation is similar or relatively shallower than those of the NLR size-luminosity relation shown as above.  

As discussed in \citet{Kang2018}, the kinematic outflow size-luminosity relation is physically different from the NLR size-luminosity 
relation. The measurement of outflow size is based on the kinematics properties of the ionized gas, which is traced by the spatially 
resolved \OIII\ kinematics. While the NLR size is measured from the distribution of photoionization, which is quantified from the 
spatial distribution of the \OIII\ emission. Both outflow and photoionization processes can transfer energy from the central 
AGN to the host galaxy. However, the outflow process can be affected by the gas distribution and interaction as well as the gravitational 
potential of the host galaxy, while the photoionization process is mainly influenced by the gas distribution. Thus, the difference of the slopes 
between the outflow size-luminosity relation and NLR size-luminosity relation presumably indicate that the outflow and photoionization 
processes have different efficiencies in transferring the AGN energy. The slop of the kinetic outflow size - luminosity relation is generally 
shallower than that of the photoionization size -luminosity relation, implying that the kinetic energy transfer is much less efficient
due to various effects including galaxy gravitational potential, local density and distribution of ISM, and the coupling efficiency between 
radiation and gas. For given AGN luminosity, ionizing photon can reach out a larger scale while kinetic energy is less effectively transferred, 
leading to a smaller kinetic outflow size than the photoionization size. We emphasize that it is important to use a proper outflow size based 
on kinematical information when outflow characteristics are constrained. If the photoionization size is used as a proxy of the outflow size, 
outflow energetics will be overestimated. Thus, it is necessary to incorporate gas kinematics in order to properly determine outflow size and 
outflow energetics.

Based on the \OIII\ line width and surface brightness of the high-velocity gas, \citet{Sun2017} define the 
kinematic disturbed region (KDR) and use it to quantify the outflow properties. They find that the measured KDR size is 
generally lower than the NLR size, which may also support the different efficiencies between outflow and photoionization 
processes in transferring the AGN energy. While they obtained a KDR size-luminosity relation with a slope of 0.6, 
which is steeper than that of our kinematic outflow size-luminosity relation. This could be due to the difference in analysis methods:
(1) they adopt a different indicator of AGN luminosity (15$\mu m$ luminosity); (2) they defined the KDR size based on an arbitrary velocity cut of 
600 km s$^{-1}$ as well as surface brightness limit, which are clearly different from our definition of the kinetic outflow size. The comparison of 
our results with those of \citet{Sun2017} suggests that the slope is sensitive to how the outflow size is defined. As pointed out by \citet{Harrison2018}, 
there are various definitions of outflow size, and a caution is required to compare literature results.

\begin{figure*}
	\centering
	\includegraphics[width=0.75\linewidth]{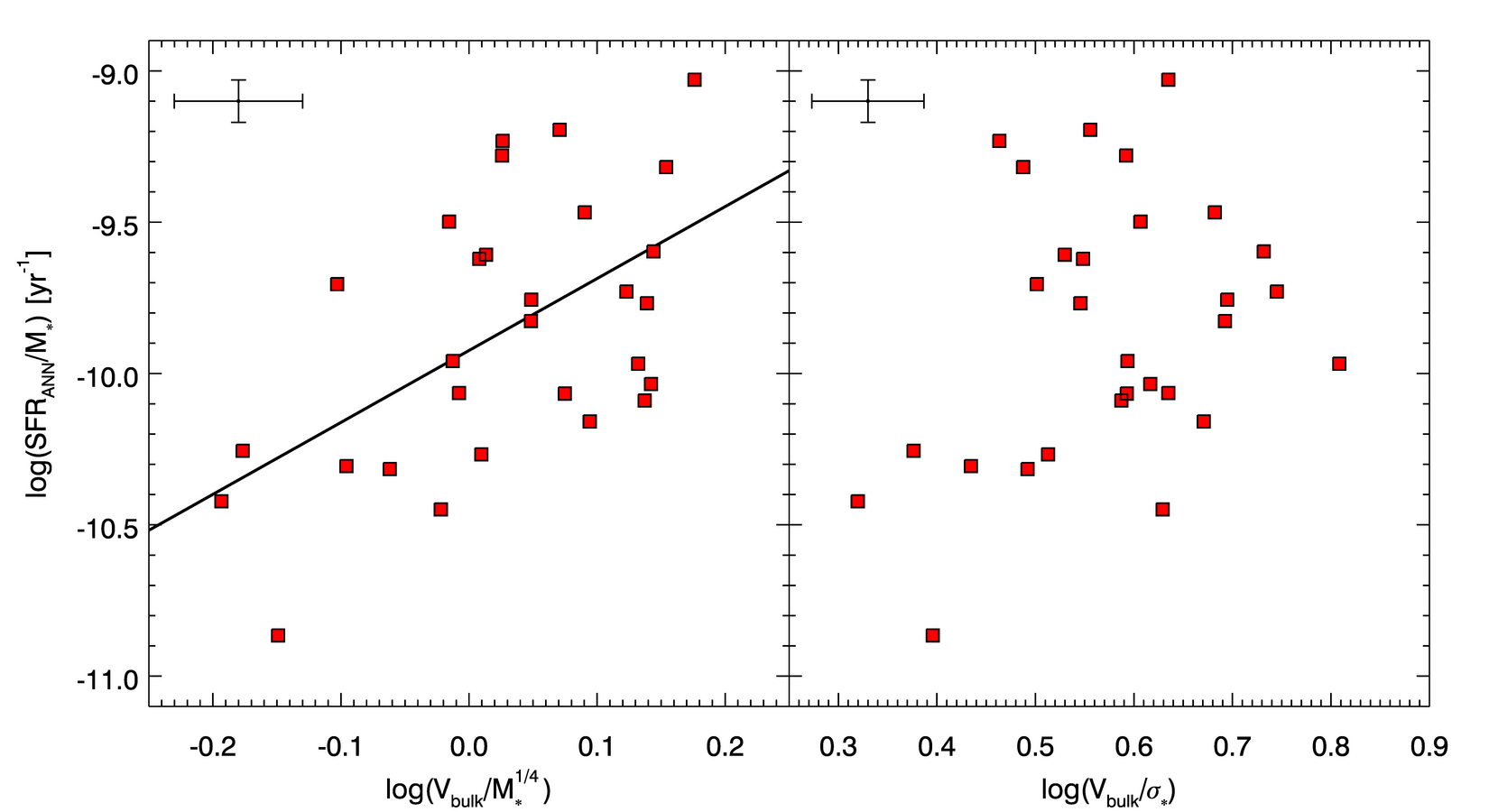}
	\caption{Comparison of the sSFR and outflow kinematics normalized by stellar mass (left) and stellar velocity dispersion (right). 
		The black line indicates the best-fit relation with a power-law index of 2.38. The typical error bars are shown in the upper left corners.}
	\label{fig:sfr1}
\end{figure*}

\subsection{Relation between Outflow Kinematics and Host Galaxy Properties}
We explore the feedback effect of AGN-driven outflows by comparing the outflow kinematics with the host galaxy properties, 
including the specific star formation rate (sSFR), D$_{n}$(4000), H$_{\delta_{A}}$, and HI gas fraction. The direct measurement 
of SFR in AGN host galaxies is difficult, since the emission from AGNs contaminate the SFR tracers, i.e., the H$\alpha$ emission 
line, UV continuum, etc (e.g., \citealt{Matsuoka2015}). We adopted the SFR from \citet{Ellison2016}, which is based on artificial 
neural network (ANN) predictions of total infrared luminosities. Then we calculate the sSFR by dividing the SFR by stellar mass from the 
MPA-JHU catalog. The D$_{n}$(4000) and H$_{\delta_{A}}$ are 
also adopted from the MPA-JHU catalog. The HI gas fraction can be estimated by adopting the photometric technique 
(e.g. \citealt{Kannappan2004,Eckert2015}), which uses a broad-band color as a proxy for the cold gas mass fraction. We use 
the $g-i$ color from SDSS photometry for estimation and adopt the calibration by \citet{Eckert2015}, as we previously applied 
(\begin{math}log(M_{HI}/M_{*})=-0.984(2.444(g-i)+0.550(b/a))+1.881\end{math}, see for details, \citealt{Luo2019}).
To quantify the strength of outflow kinematics, we first estimate the bulk velocity of the outflowing gas by adding the \OIII\ velocity 
shift and dispersion of each spaxel in quadrature and calculate the flux-weighted mean within the kinematic outflow size. 
Then we normalize it with the stellar mass or stellar velocity dispersion which are considered as global indicators of the 
gravitational potential of the host galaxy. The stellar velocity dispersion is also adopted from the MPA-JHU catalog.
We summarise the host galaxy and outflow properties in Table \ref{tab:properties}.

\begin{figure}
	\centering
	\includegraphics[width=1.0\linewidth]{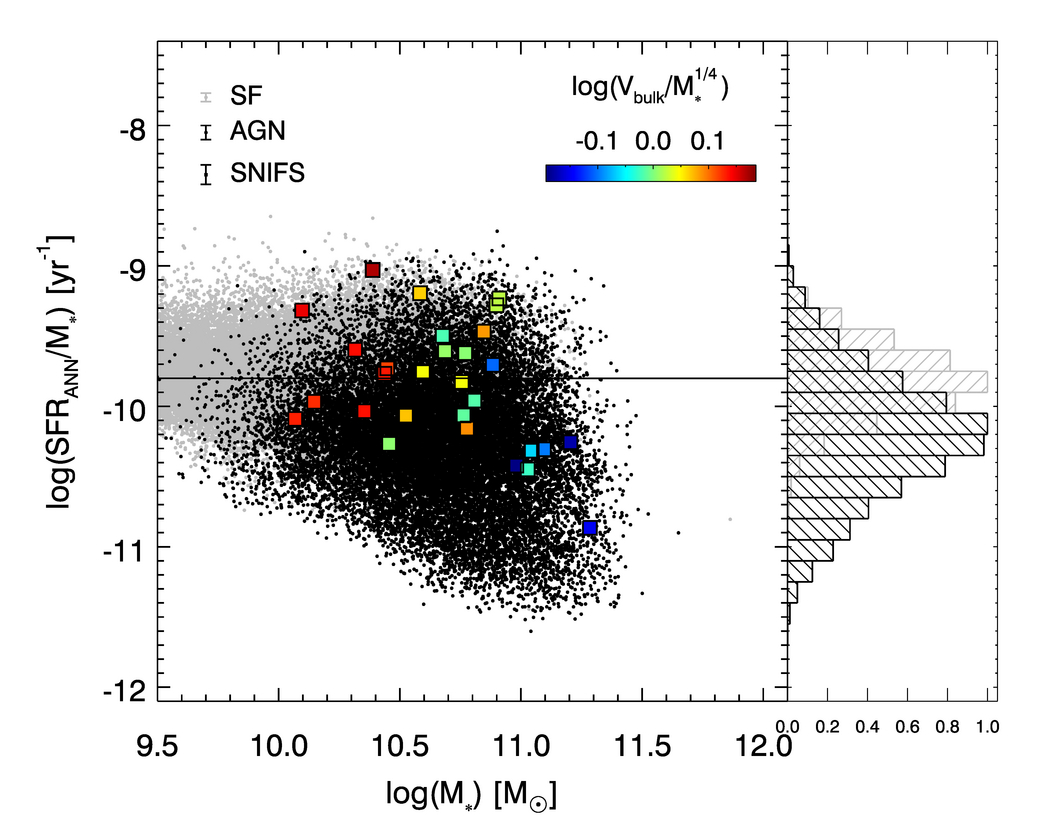}
	\caption{The relation between the stellar mass and sSFR. Two large samples of star forming galaxies and AGNs at z $<$ 0.3 are 
		shown in gray and black dots, respectively. The black horizontal line indicates the median log(sSFR) of the star forming 
		galaxies in the main sequence. The typical error bars are shown in the upper left corner. 
		Different colors represent the outflow kinematics normalized by stellar mass.}
	\label{fig:sfr2}
\end{figure}

\begin{figure}
	\centering
	\includegraphics[width=0.85\linewidth]{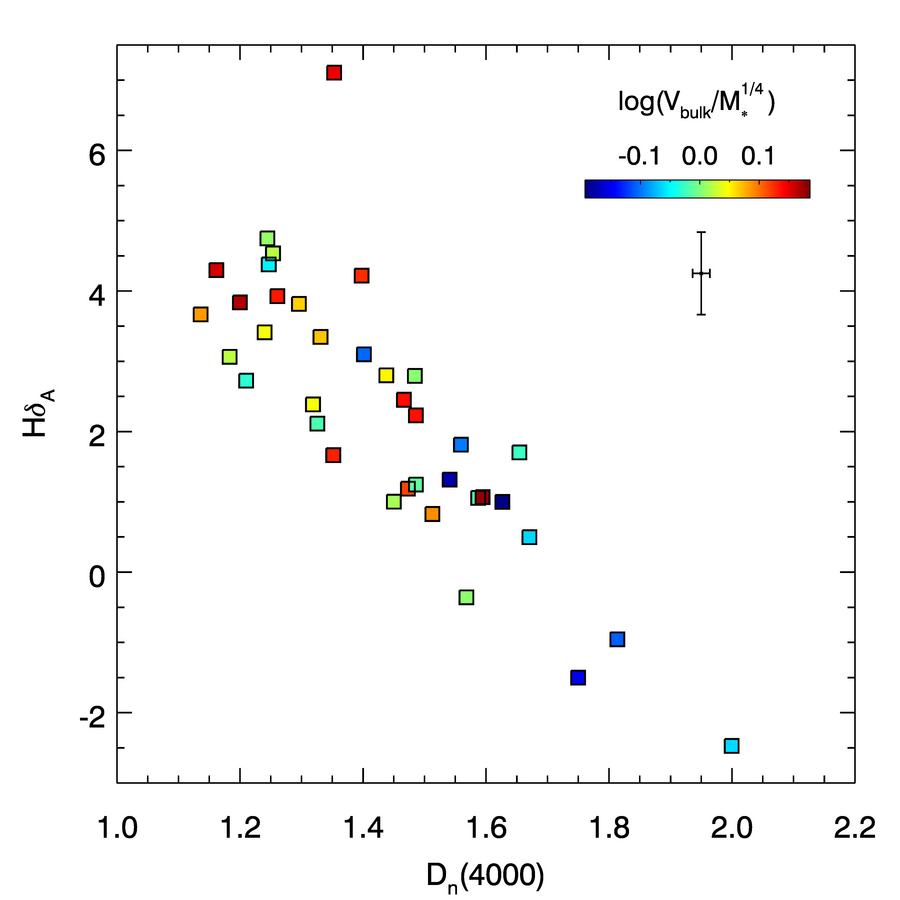}
	\caption{The D$_{n}$(4000) vs. H$\delta_{A}$ plane. The typical error bars are shown in the upper right corner. 
		Different colors represent the outflow kinematics normalized by stellar mass.}
	\label{fig:d4000}
\end{figure}

As shown in Figure \ref{fig:sfr1}, the sSFR of the host galaxy is positively correlated with the bulk velocity normalized by stellar mass, 
which indicates that AGNs with strong outflows tend to have a higher sSFR than that of the AGNs with weak outflows. The Spearman’s 
rank correlation coefficient is 0.5, with a p-value $< 5.6\times10^{-3}$. By removing the effect of stellar mass, the partial correlation 
coefficient is 0.3. The best-fitted relation between the sSFR and bulk velocity normalized by stellar mass can be quantified as a single 
power law with an index of 2.38. While for the bulk velocity normalized by the stellar velocity dispersion, we can not find a 
clear correlation with the sSFR of the host galaxy. Although the correlation is not strong, it seems that there is a broad trend between 
sSFR and the normalized bulk velocity by stellar velocity dispersion. Since stellar velocity dispersion suffers various effects, including 
inclination to the line-of-sight, and aperture size, stellar mass may better represent the host galaxy gravitational potential in normalizing 
gas bulk velocity. In Figure \ref{fig:sfr2}, we compare our AGN sample with large samples of star-forming galaxies and AGNs in the 
diagram of sSFR versus stellar mass. The large samples of star-forming galaxies and AGNs are adopted from \citep{Woo2017}, which 
are selected from the emission line galaxies at z $<$ 0.3, based on the archival spectra of the SDSS Data Release 7. The solid line 
indicates the median log(sSFR) of the star-forming galaxies in the main sequence. The AGNs with strong outflows have comparable 
sSFR as the star-forming galaxies in the main sequence, while the average sSFR of the AGNs with weak outflows is one dex lower. 
We also present the diagram of D$_{n}$(4000) versus H$_{\delta_{A}}$ in Figure \ref{fig:d4000}, which is color-coded by the 
normalized bulk velocity of the outflowing gas. The AGNs with strong outflows have relatively lower D$_{n}$(4000) and higher 
H$_{\delta_{A}}$ than those with weak outflows. Since D$_{n}$(4000) and H$_{\delta_{A}}$ are indicators of SFR, this result 
implies more intensive star formation in the host galaxies of AGNs with strong outflows, in agreement with the above results 
obtained from the IR-based SFR. 

The relations between outflow kinematics and star formation are generally consistent with our previous statistical study of 
outflow impact in a large sample of Type-1 and Type-2 AGNs \citep{Woo2017,Woo2020}, which showed that star formation rate 
is higher for strong outflow AGNs, suggesting no instantaneous suppression or quenching of star formation while the effect of 
feedback is delayed. When the AGNs and star formation are triggered by enough gas supply, the AGNs are strong and can produce 
strong outflows, while the star formation is on-going as in regular star forming galaxies. If the outflows can only impact the ISM 
after a certain timescale, although the feedback effect can be observed as the SFR decrease, the AGNs will also become weak and 
only show weak outflows. While our study focus on AGNs in the local universe, several studies have also shown that AGNs at 
higher redshift (e.g. z$\sim$2 ) do not have instantaneous impact on the star formation in the host galaxies \citep{Harrison2017}. 
\citet{Scholtz2020} examine the spatial anti-correlation between the AGN-driven outflows and in-situ star formation in eight 
moderate-luminosity AGNs at z$=$1.4–2.6. In three targets with significant outflow signatures, they can not find a clear evidence 
that outflows suppress the star formation. \citet{Davies2020} find powerfull AGN-driven outflows ($V_{out} \sim$ 1500 km s$^{-1}$) 
in two compact star forming galaxies at z$=$2.2, which are located at the upper envelope of the star-forming main sequence, 
suggesting no instantaneous feedback effect of AGN in these galaxies.

\begin{figure}
	\centering
	\includegraphics[width=0.85\linewidth]{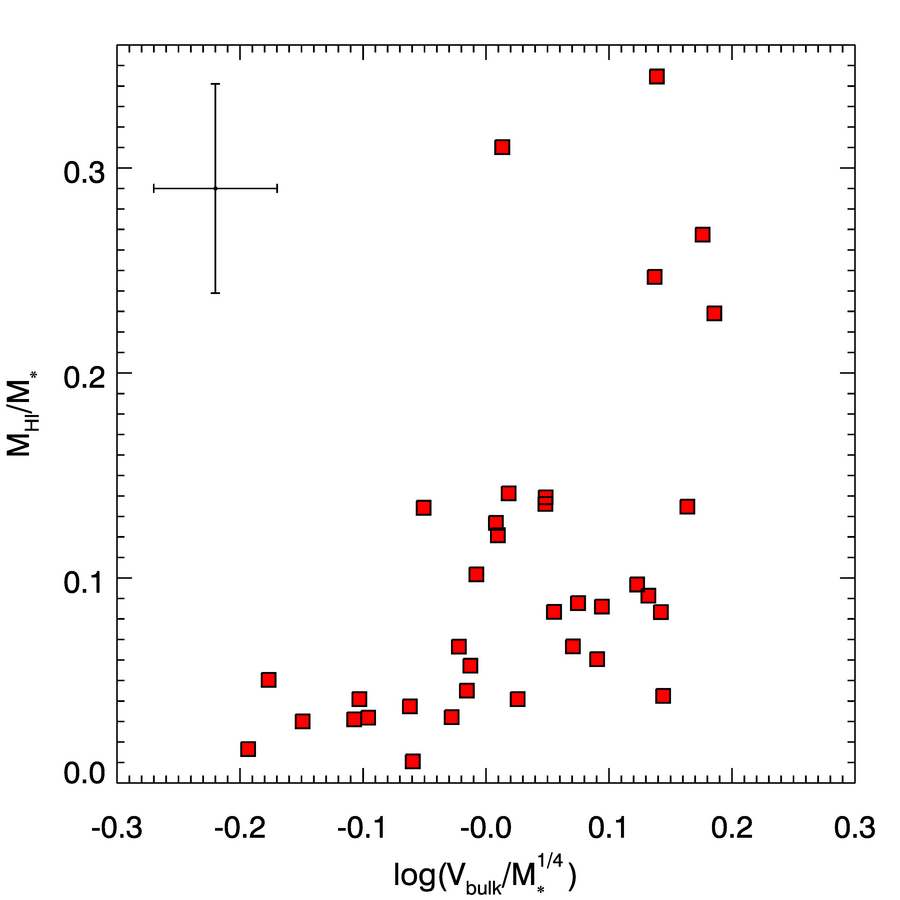}
	\caption{Comparison of the HI gas fraction and outflow kinematics normalized by stellar mass. 
		The typical error bars are shown in the upper left corner.}
	\label{fig:gasf}
\end{figure}

As discussed in \citet{Woo2017,Woo2020}, the aforementioned link between AGN outflows and sSFR 
can also be explained as a natural consequence of gas fraction decreasing due to the internal or external mechanisms. 
When there is a large amount of gas in the host galaxy, both AGN and star formation are strong, which enables 
us to observe the strong outflows. As the supplied gas is depleted, the AGN activity and star formation will 
decrease, and we can only observe the weak outflows in the host galaxies with lower SFR. In addition, it is 
also possible that the gas fraction is intrinsically different among the host galaxies. We examine the relation 
between the strength of outflow kinematics and HI gas fraction in Figure \ref{fig:gasf}. For the AGNs with 
strong outflows, the HI gas fraction tends to increase with the normalized bulk velocity and can reach the 
relatively high value of 0.3, while the HI gas fraction is on average lower in the AGNs with weak outflows 
and does not vary with the normalized bulk velocity. This result indicates the importance of cold gas content 
in understanding the relation between AGN-driven outflows and sSFR. Since we can not trace the time evolution 
of the intrinsic gas fraction, it is difficult to test the above scenarios. Cosmological simulations like IllustrisTNG 
may help to investigate the time evolution of the gas content in the host galaxies of AGNs and provide hints 
on the exact scenario. The relation between HI gas fraction and normalized bulk velocity also suggests that 
the presence of outflows may depend on the gas fraction, which has been discussed in \citet{Luo2019} based 
on a large sample of AGNs with and without outflow. As the HI gas fraction is estimated indirectly, 
multiwavelength follow-up observations will provide better constraints on the intrinsic gas fraction 
and enable us to further study the connection between cold gas, star formation, and AGN outflows.

\section{Summary}
\label{sec:summary}
To investigate the feedback effect of AGN-driven outflows, we perform integral-field spectroscopic observations of 40 moderate-luminosity 
($10^{41.5} < L_{\mathrm{\OIII;cor}} < 10^{43.1}$ erg s$^{-1}$ ) Type-2 AGNs at z $<$ 0.1. These targets are selected from the a large sample 
of $\sim$39,000 Type-2 AGNs based on their integrated \OIII\ kinematics, which provide us a uniform sample to quantify the outflow 
properties and compare them with the properties of AGN and host galaxy. We summarize the main results as below.

\begin{itemize}
\item Based on the radial profile of the normalized \OIII\ velocity dispersion by stellar velocity dispersion, we measure the kinematic 
outflow size, which ranges from 0.9 to 4 kpc with a median value of 2 kpc. By including the size measurements from \citet{Kang2018}, 
we confirm that the kinematic outflow size is well correlated with the \OIII\ luminosity. The slope of the best-fitted relation is $0.23\pm0.02$, 
which is slightly lower than that obtained in \citet{Kang2018}. Our results further confirm the kinematic outflow size-luminosity relation 
reported in \citet{Kang2018} and extend it into a broader \OIII\ luminosity range (over four orders of magnitude in [O III] luminosity).

\item We compare the outflow kinematics with the host galaxy properties, including the sSFR, D$_{n}$(4000), H$_{\delta_{A}}$, and 
HI gas fraction. We find a clear correlation between the sSFR and normalized outflow velocity. The AGNs with strong outflows have 
comparable sSFR to the star-forming galaxies in the main sequence, while the AGNs with weak outflows present much lower sSFR. 
In addition, we also find that AGNs with strong outflows have higher star formation rate and HI gas fraction than those with weak 
outflows. These results are consistent with our previous studies of outflow impact in large samples of AGNs \citep{Woo2017,Woo2020}, 
suggesting that the current feedback from AGN-driven outflows do not instantaneously suppress or quench the star formation 
in the host galaxies while its effect is delayed. 

\end{itemize}

\acknowledgments{We thank the referee for various comments. We thank Hai Fu for his help on the observation 
	and reduction of SNIFS data. This research was supported by the 
	National Research Foundation of Korea (NRF) grant funded by the Korea government (MEST) (No. 2016R1A2B3011457). 
	J.H.W thanks the Korea Astronomy and Space Science Institute for its hospitality during a sabbatical visit.This work is based 
	on the observations made with the University of Hawaii 88-inch Telescope (UH88). SNIFS on the UH88 telescope is part of the Nearby 
	Supernova Factory II project, a scientific collaboration among the Centre de Recherche Astronomique de Lyon, Institut de Physique 
	Nucl\^{e}aire de Lyon, Laboratoire de Physique Nucl\^{e}eaire et des Hautes Energies, Lawrence Berkeley National Laboratory, 
	Yale University, University of Bonn, Max Planck Institute for Astrophysics, Tsinghua Center for Astrophysics, and the Centre de 
	Physique des Particules de Marseille.}

\bibliographystyle{aasjournal}
\bibliography{snifs_paper_accepted_arXiv}

\appendix
\renewcommand\thefigure{\thesection.\arabic{figure}}
\section{Additional Figures}

\setcounter{figure}{0}

\begin{figure*}
	\begin{adjustbox}{addcode={\begin{minipage}{\width}}{\caption{
			}\end{minipage}},rotate=90,center}
		\includegraphics[width=1.25\linewidth]{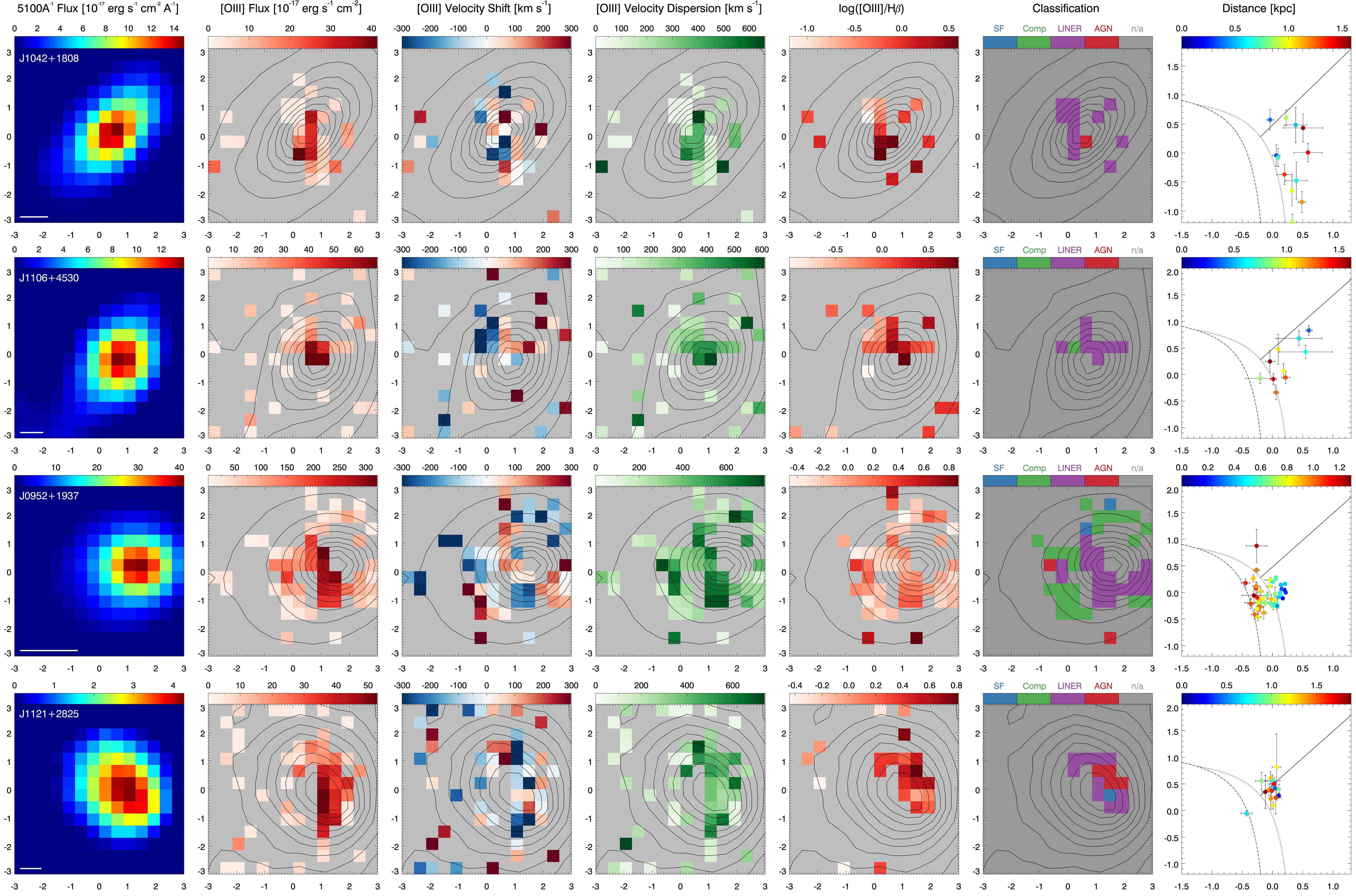}
	\end{adjustbox}
\end{figure*}

\begin{figure*}
	\begin{adjustbox}{addcode={\begin{minipage}{\width}}{\caption{
			}\end{minipage}},rotate=90,center}
		\includegraphics[width=1.25\linewidth]{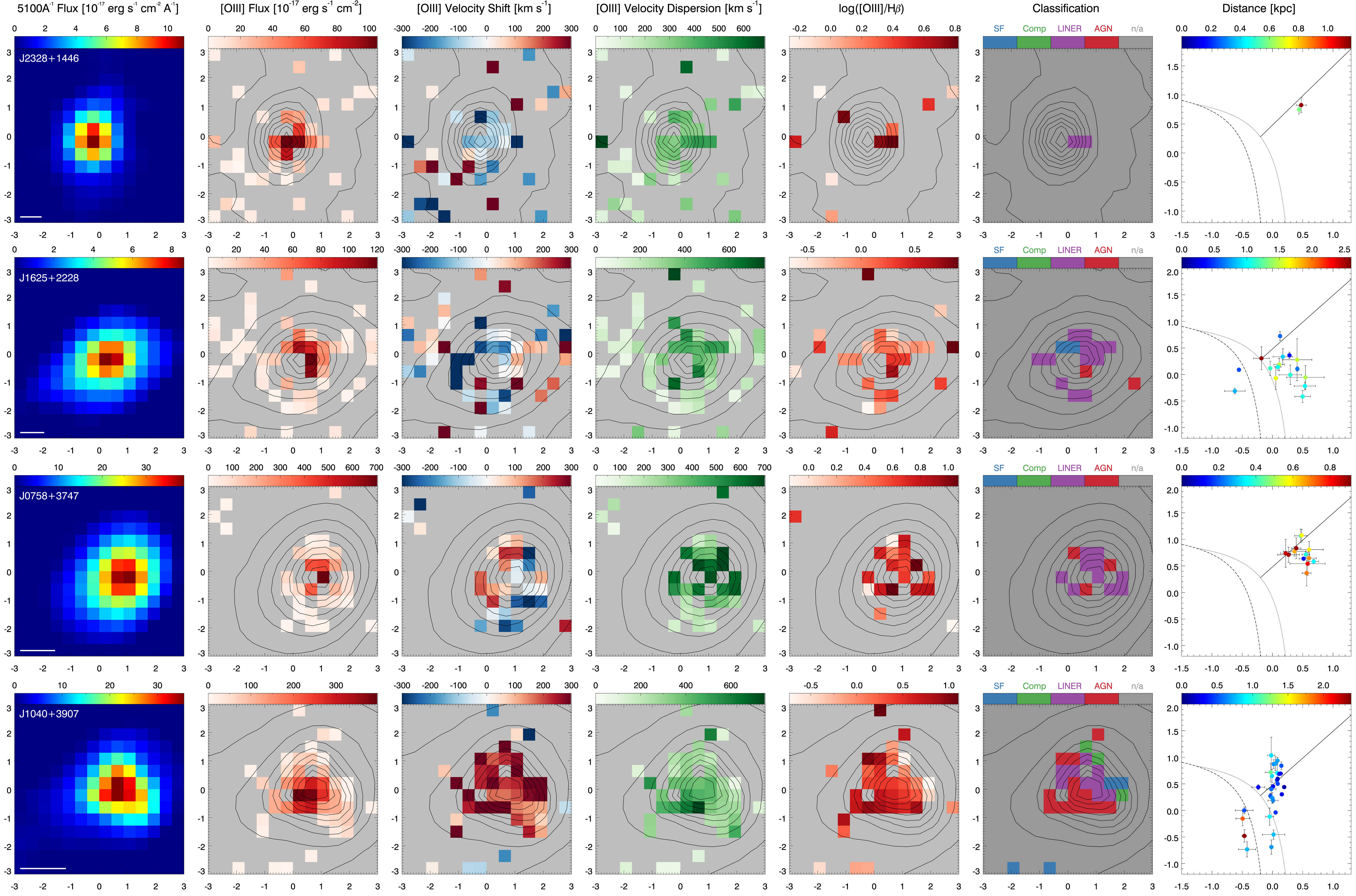}
	\end{adjustbox}
\end{figure*}

\begin{figure*}
	\begin{adjustbox}{addcode={\begin{minipage}{\width}}{\caption{
			}\end{minipage}},rotate=90,center}
		\includegraphics[width=1.25\linewidth]{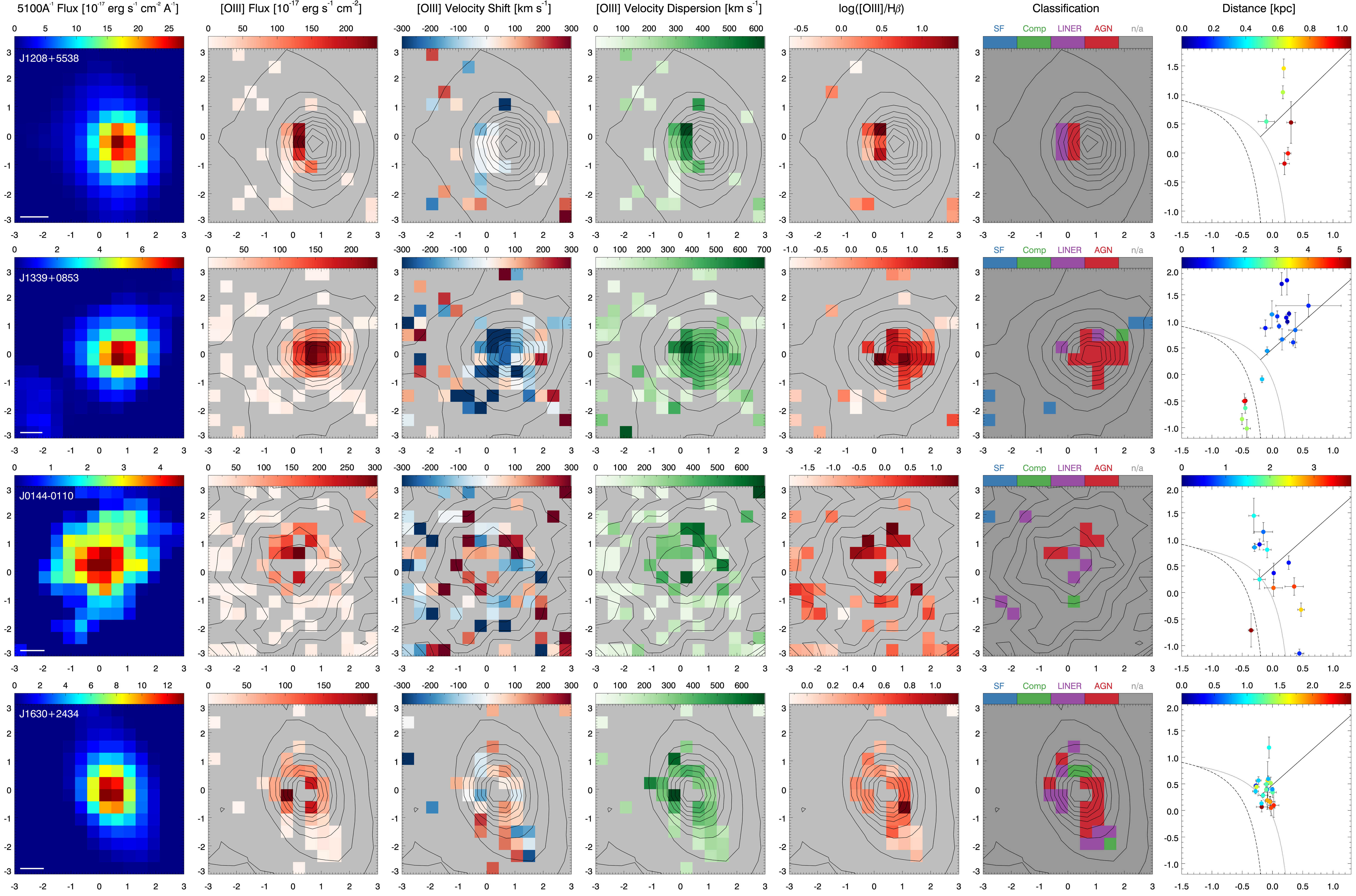}
	\end{adjustbox}
\end{figure*}

\begin{figure*}
	\begin{adjustbox}{addcode={\begin{minipage}{\width}}{\caption{
			}\end{minipage}},rotate=90,center}
		\includegraphics[width=1.25\linewidth]{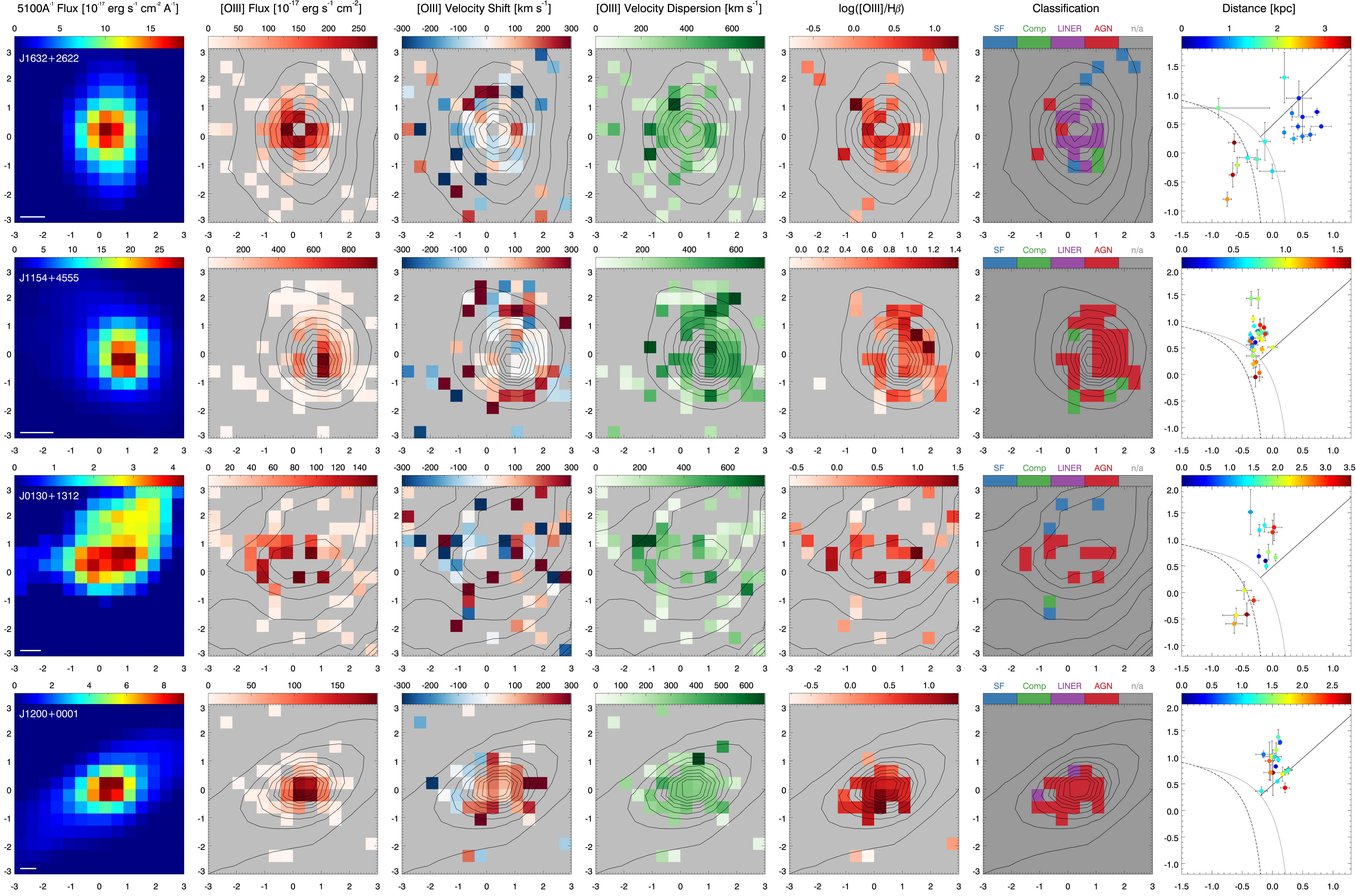}
	\end{adjustbox}
\end{figure*}

\begin{figure*}
	\begin{adjustbox}{addcode={\begin{minipage}{\width}}{\caption{
			}\end{minipage}},rotate=90,center}
		\includegraphics[width=1.25\linewidth]{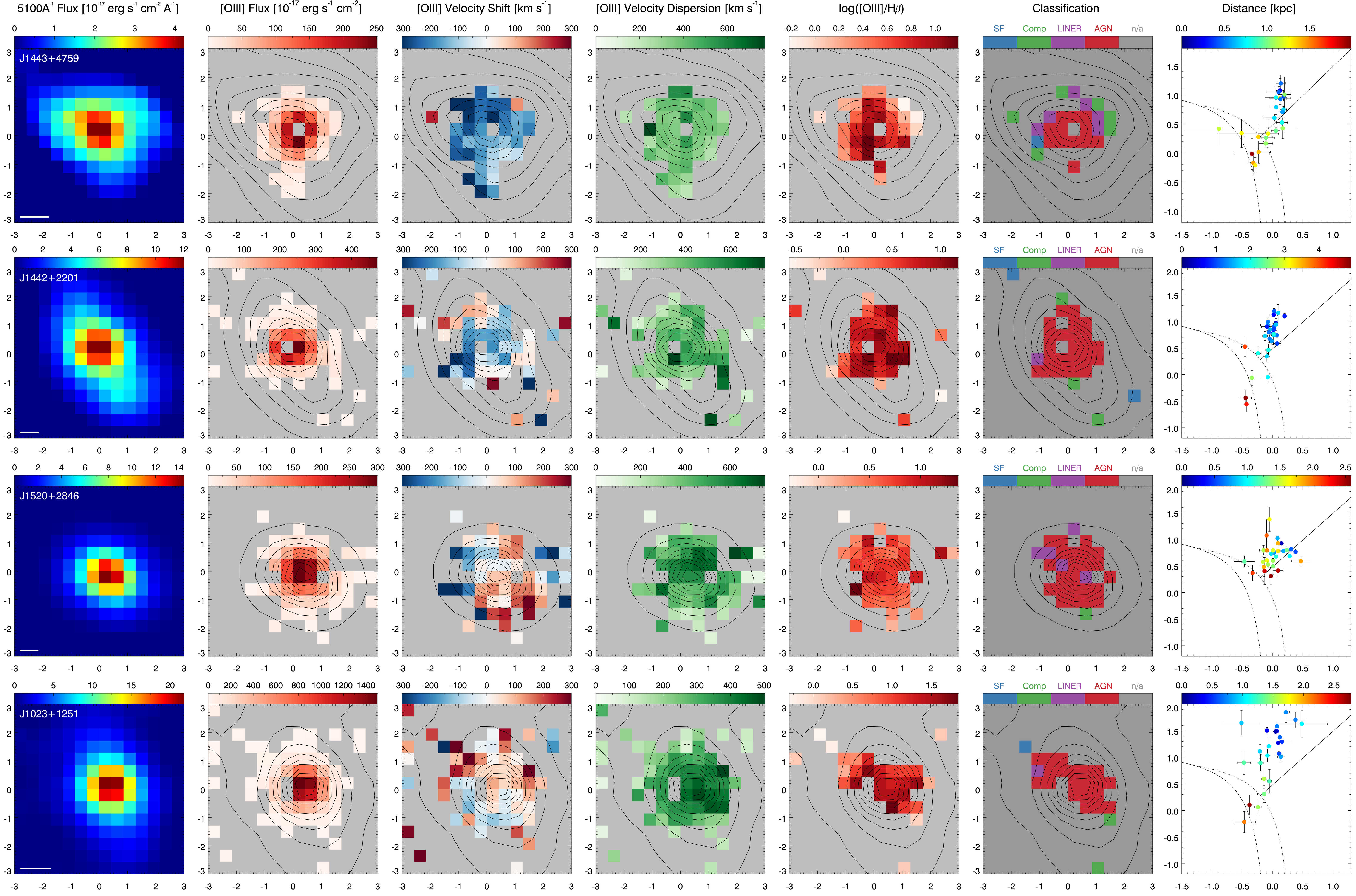}
	\end{adjustbox}
\end{figure*}

\begin{figure*}
	\begin{adjustbox}{addcode={\begin{minipage}{\width}}{\caption{
			}\end{minipage}},rotate=90,center}
		\includegraphics[width=1.25\linewidth]{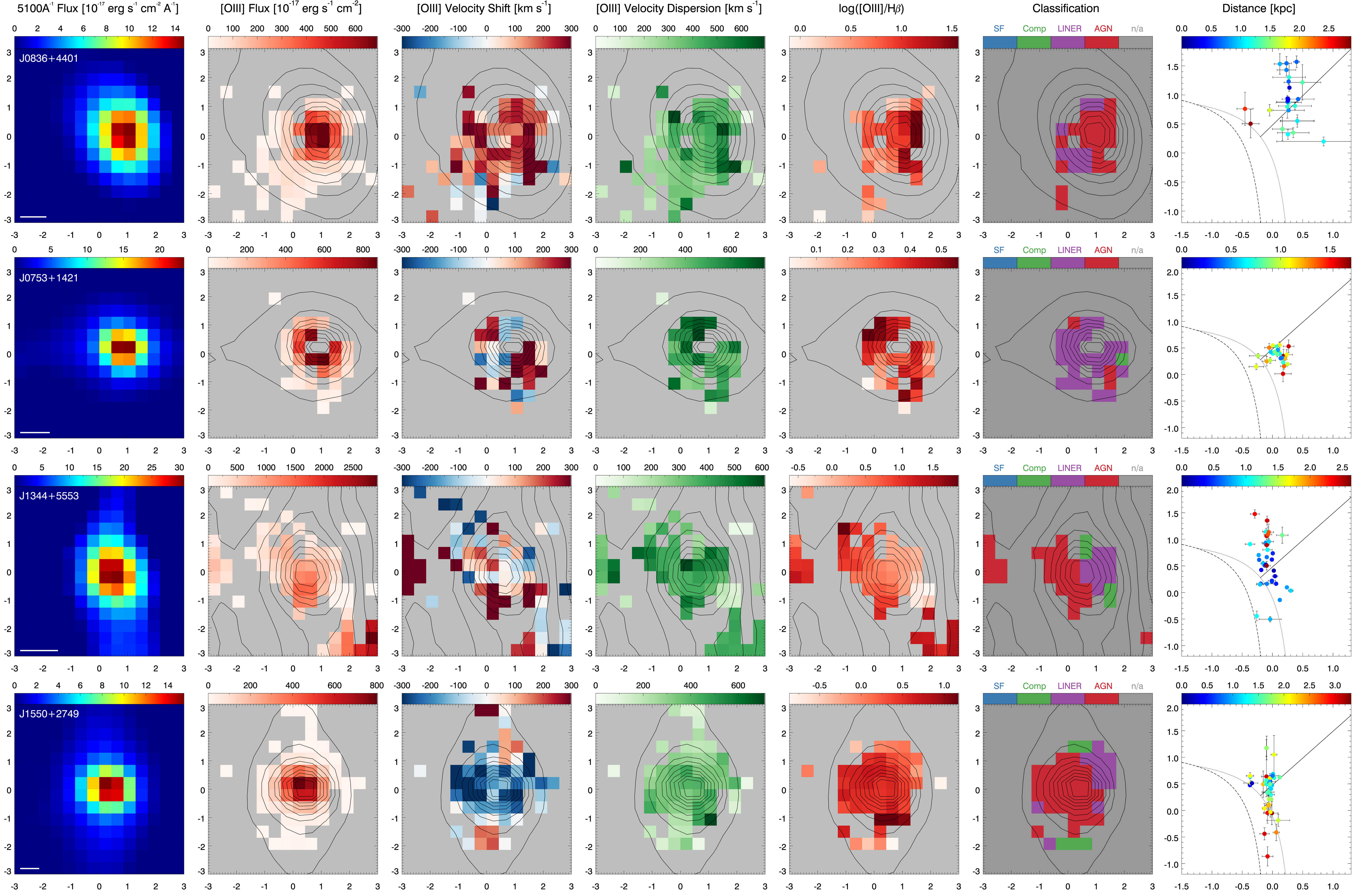}
	\end{adjustbox}
\end{figure*}

\begin{figure*}
	\begin{adjustbox}{addcode={\begin{minipage}{\width}}{\caption{
			}\end{minipage}},rotate=90,center}
		\includegraphics[width=1.25\linewidth]{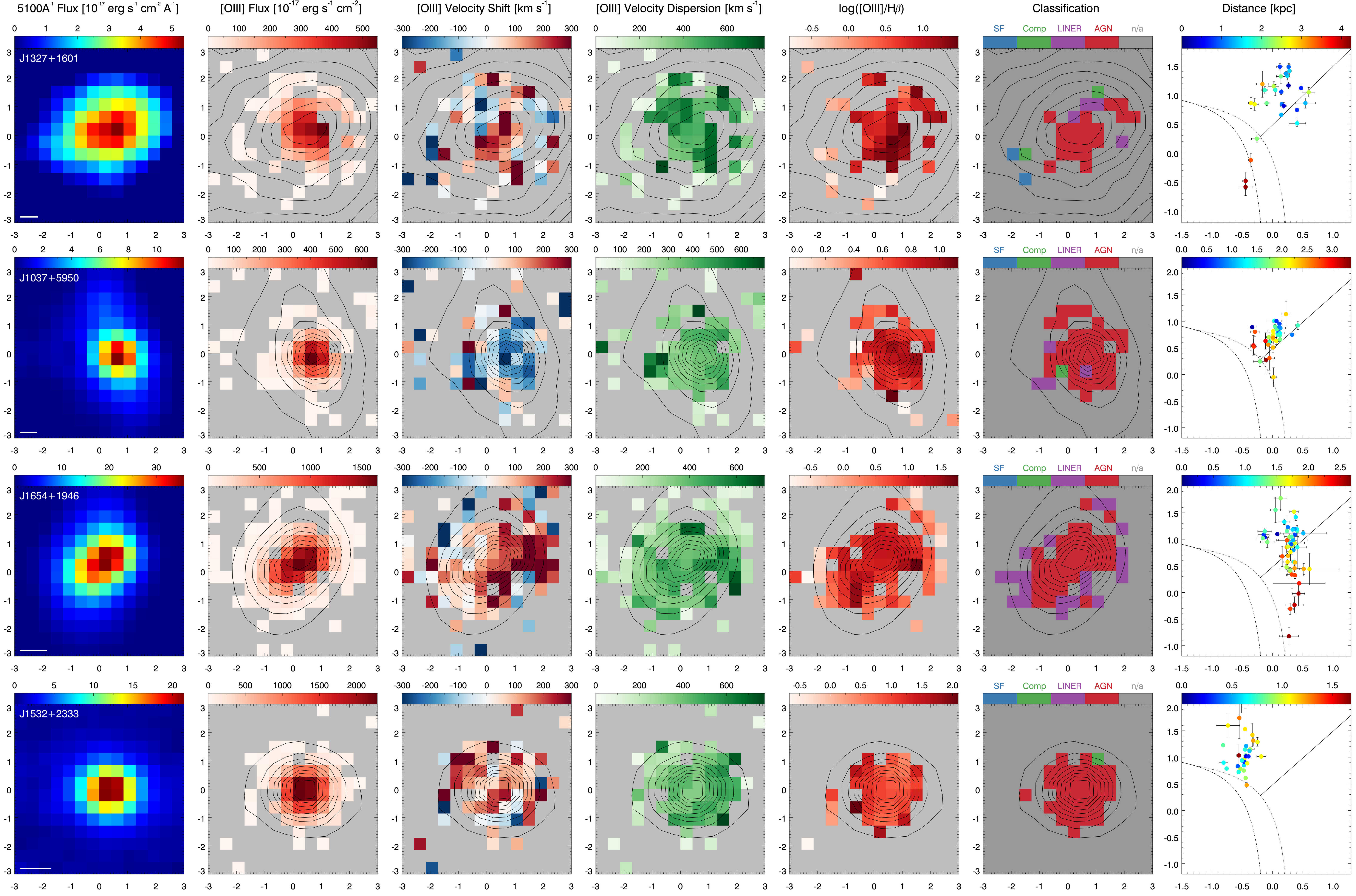}
	\end{adjustbox}
\end{figure*}

\begin{figure*}
	\begin{adjustbox}{addcode={\begin{minipage}{\width}}{\caption{
			}\end{minipage}},rotate=90,center}
		\includegraphics[width=1.25\linewidth]{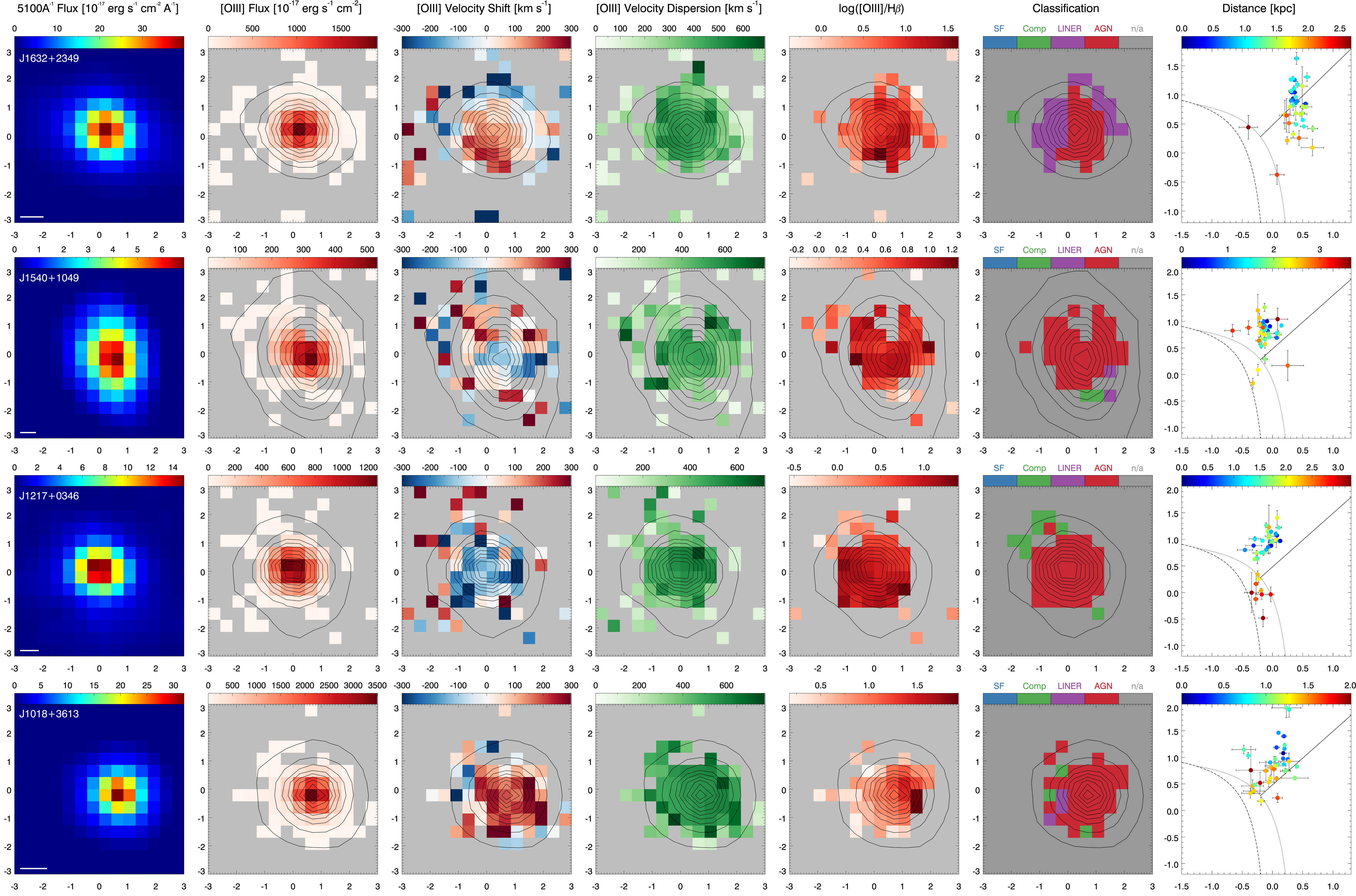}
	\end{adjustbox}
\end{figure*}

\end{document}